\documentclass[twocolumn, prl, 10pt, superscriptaddress]{revtex4-1}
\usepackage{graphicx, subfigure, amsfonts,amssymb,amsmath, array, gensymb,fontenc,color, lipsum}
\usepackage{booktabs} 
\newcommand{\ra}[1]{\renewcommand{\arraystretch}{#1}}
\usepackage{booktabs}
\usepackage{multirow}
\begin{document}

\title{Electronic excitations and spin interactions in chromium trihalides \\ from embedded many-body wavefunctions}

\author{Ravi Yadav}
\affiliation{Institute of Physics, Ecole Polytechnique F\'ed\'erale de Lausanne (EPFL), CH-1015 Lausanne, Switzerland}
\affiliation{National Centre for Computational Design and Discovery of Novel Materials (MARVEL),  Ecole Polytechnique F\'ed\'erale de Lausanne (EPFL), CH-1015 Lausanne, Switzerland}

\author{Lei Xu}
 \thanks{Present Address: Theoretical Division, Los Alamos National Laboratory, Los Alamos, New Mexico 87544, USA}
\affiliation {Institute for Theoretical Solid State Physics, IFW Dresden, Helmholtzstr.~20, 01069 Dresden, Germany}

\author{Michele Pizzochero}
\affiliation{School of Engineering and Applied Sciences, Harvard University, Cambridge, MA 02138, United States}

\author{Jeroen van den Brink}
\affiliation {Institute for Theoretical Solid State Physics, IFW Dresden, Helmholtzstr.~20, 01069 Dresden, Germany}
\affiliation{Institute for Theoretical Physics and W\"urzburg-Dresden Cluster of Excellence ct.qmat, Technische Universit\"at Dresden, 01069 Dresden, Germany}


\author{Mikhail I. Katsnelson}
\affiliation{Institute for Molecules and Materials, Radboud University, 6525AJ Nijmegen, The Netherlands}


\author{Oleg V.\ Yazyev}
\email{Electronic address: \url{oleg.yazyev@epfl.ch}}
\affiliation{Institute of Physics, Ecole Polytechnique F\'ed\'erale de Lausanne (EPFL), CH-1015 Lausanne, Switzerland}
\affiliation{National Centre for Computational Design and Discovery of Novel Materials (MARVEL),  Ecole Polytechnique F\'ed\'erale de Lausanne (EPFL), CH-1015 Lausanne, Switzerland}

\date{\today}
\begin{abstract}
Although chromium trihalides are widely regarded as a promising class of two-dimensional magnets for next-generation devices, an accurate description of their electronic structure and magnetic interactions has proven challenging to achieve. Here, we quantify electronic excitations and spin interactions in Cr$X_3$ ($X=$~Cl, Br, I) using embedded many-body wavefunction calculations and fully generalized spin Hamiltonians. We find that the three trihalides feature comparable $d$-shell excitations, consisting of a high-spin $^4A_2$ $(t^3_{2g}e^0_{g})$ ground state lying 1.5$-$1.7 eV below the first excited state  $^4T_2$ ($t^2_{2g}e^1_{g}$). CrCl$_3$ exhibits a single-ion anisotropy $A_{\rm sia} = -0.02$ meV, while the Cr spin-3/2 moments are ferromagnetically coupled through bilinear and biquadratic exchange interactions of $J_1 = -0.97$ meV and $J_2 = -0.05$ meV, respectively. The corresponding values for CrBr$_3$ and CrI$_3$ increase to $A_{\rm sia} = -0.08$ meV and $A_{\rm sia} = -0.12$ meV for the single-ion anisotropy,  $J_1 = -1.21$ meV, $J_2 = -0.05$ meV and $J_1 = -1.38$ meV, $J_2 = -0.06$ meV for the exchange couplings, respectively. We find that the overall magnetic anisotropy is defined by the interplay between $A_{\rm sia}$ and $A_{\rm dip}$ due to magnetic dipole-dipole interaction that favors in-plane orientation of magnetic moments in ferromagnetic monolayers and bulk layered magnets. The competition between the two contributions sets CrCl$_3$ and CrI$_3$ as the easy-plane ($A_{\rm sia}+ A_{\rm dip} > 0$) and easy-axis ($A_{\rm sia}+ A_{\rm dip} < 0$) ferromagnets, respectively. The differences between the magnets trace back to the atomic radii of the halogen ligands and the magnitude of spin-orbit coupling. Our findings are in excellent agreement with recent experiments, thus providing reference values for the fundamental interactions in chromium trihalides.
\end{abstract}

\maketitle

The advent of two-dimensional (2D) magnets has created new opportunities to explore and manipulate spin interactions in the ultimate limit of atomic thickness, holding promise for an array of nanoscale applications ranging from magneto-optoelectronics to quantum information  \cite{Burch2018, Gibertini2019, Gong2019, Mak2019}. In these 2D magnets, the non-vanishing critical temperature originates from the magnetic anisotropy, which is key to preserving long-ranged magnetic order against thermal fluctuations, as explained by the theorems of Mermin-Wagner  \cite{Mermin1966} and Hohenberg \cite{Hohenberg1967}. Within the rapidly expanding class of ultrathin magnets, layered chromium trihalides of general formula Cr$X_3$ ($X = $ Cl, Br, I) \cite{Soriano2020} are among the most discussed members by virtue of their versatile magnetic properties that can be engineered via, e.g., atomic thickness \cite{Huang2017}, stacking configuration \cite{Sivadas2018, Song2019}, electrostatic gating \cite{Huang2018, Jiang2018}, lattice deformations \cite{Tingxin2019, Pizzochero2020a,PhysRevB.98.144411}, defects incorporation \cite{Wang2020, Pizzochero2020b}, or light irradiation \cite{Singamaneni2020}. 
Such tunability is key for the construction of new energy-efﬁcient spin-based devices.

In the recent years, a flurry of theoretical investigations on this family of materials established the important role played by $t_{2g}$-$e_g$ interactions\,\cite{PhysRevMaterials.3.031001}, ligand $p$ orbitals\,\cite{PhysRevB.99.104432,soriano_environmental_2021} and stacking sequence \,\cite{Wang21} in stabilizing the ferromagnetic order. The insulating chromium trihalides realize highly localized magnetic moments close to 3 $\mu_B$, corresponding to an $S = 3/2$ system \cite{McGuire2017}, with short-range superexchange interactions, as well as local magnetic anisotropy. The latter allows this family of 2D materials to overcome the Mermin-Wagner theorem and to establish long-range magnetic order \cite{PhysRevB.60.1082}. 
Inelastic neutron scattering experiments on CrI$_3$ showed $\approx$4~meV gap opening between the two magnon branches~\cite{PhysRevX.8.041028}. Chen {\it et al.} \cite{PhysRevX.8.041028} argued that the gap originates from the Dzyaloshinskii-Moriya (DM) interaction and predicted an unusually large dominant DM interaction from the fitting of magnon spectra. However, the nearest-neighbor (NN) DM vector in CrI$_3$ is not allowed by symmetry and only the next NN DM interaction is finite. Alternatively, such a gap was argued to result from significant Kitaev interactions (symmetric anisotropic exchange)\,\cite{PhysRevLett.124.017201}. The Kitaev term is allowed by symmetry, but similar to DM interaction, it also originates from spin-orbit coupling (SOC) and is usually quite small in 3$d$ materials. The heavier ligands do contribute to SOC, but the presence of such dominant anisotropic interaction is still under debate as the magnetic interaction parameters reported from various experiments and theoretical calculations differ\,\cite{Soriano2020,PhysRevB.102.115162,PhysRevLett.124.017201,ke_electron_2021}. Further studies have also discussed higher order interactions (biquadratic exchange) and lattice defects as the origin of the gap\,\cite{Kartsev2020, Wahab2021}.   

Despite the rapidly growing interest in Cr$X_3$ magnets, an accurate description of magnetic interactions as well as electronic excitations is still needed. On the one hand, high-resolution measurements based on soft x-ray spectroscopies, which are instrumental to accessing electronic excitations, are challenging  to accomplish and tedious to interpret \cite{Shao2021}. On the other hand, recent experiments assessing spin interactions are mainly limited to isotropic Heisenberg exchange couplings, the strength of which remains debated since reported values differ by up to one order of magnitude \cite{Soriano2020, Lee2020, Cenker2021}. From the theoretical point of view, earlier studies primarily resort to density-functional methods, which are inherently inadequate to cope with both excited and correlated electronic states. Such a drawback was shown to be particularly severe when dealing with chromium trihalides and other 2D magnets, where, depending on the adopted exchange-correlation approximation and computational scheme, the strength of the calculated magnetic interactions spreads over a wide energy window \cite{Soriano2020, Pizzochero2020c, Menichetti2019}. This calls for a rigorous and unambiguous account of electron-electron effects in order to achieve a detailed comprehension of Cr$X_3$ magnets.

In this work, we present a comparative investigation of electronic excitations and spin interactions in two-dimensional Cr$X_3$ ($X = $ Cl, Br, I) using post-Hartree-Fock wavefunctions, arguably the most accurate and rigorous approach to treat electronic correlations.
First, we determine the $d$-shell multiplet structures, which allows us to offer an unambiguous interpretation to recent x-ray spectroscopic measurements. Next, we quantify the strength of dominant magnetic interactions, including single-ion anisotropy, $g$-factors, and exchange couplings, using a fully generalized spin Hamiltonian. Altogether, our findings establish a theoretical ground to recent spectroscopic observations, thus providing a quantitative understanding of the microscopic physics in chromium trihalides from a quantum chemical perspective.

\section{Methodology} 
\begin{figure*}[h!]
    \centering
    \includegraphics[width=1\columnwidth]{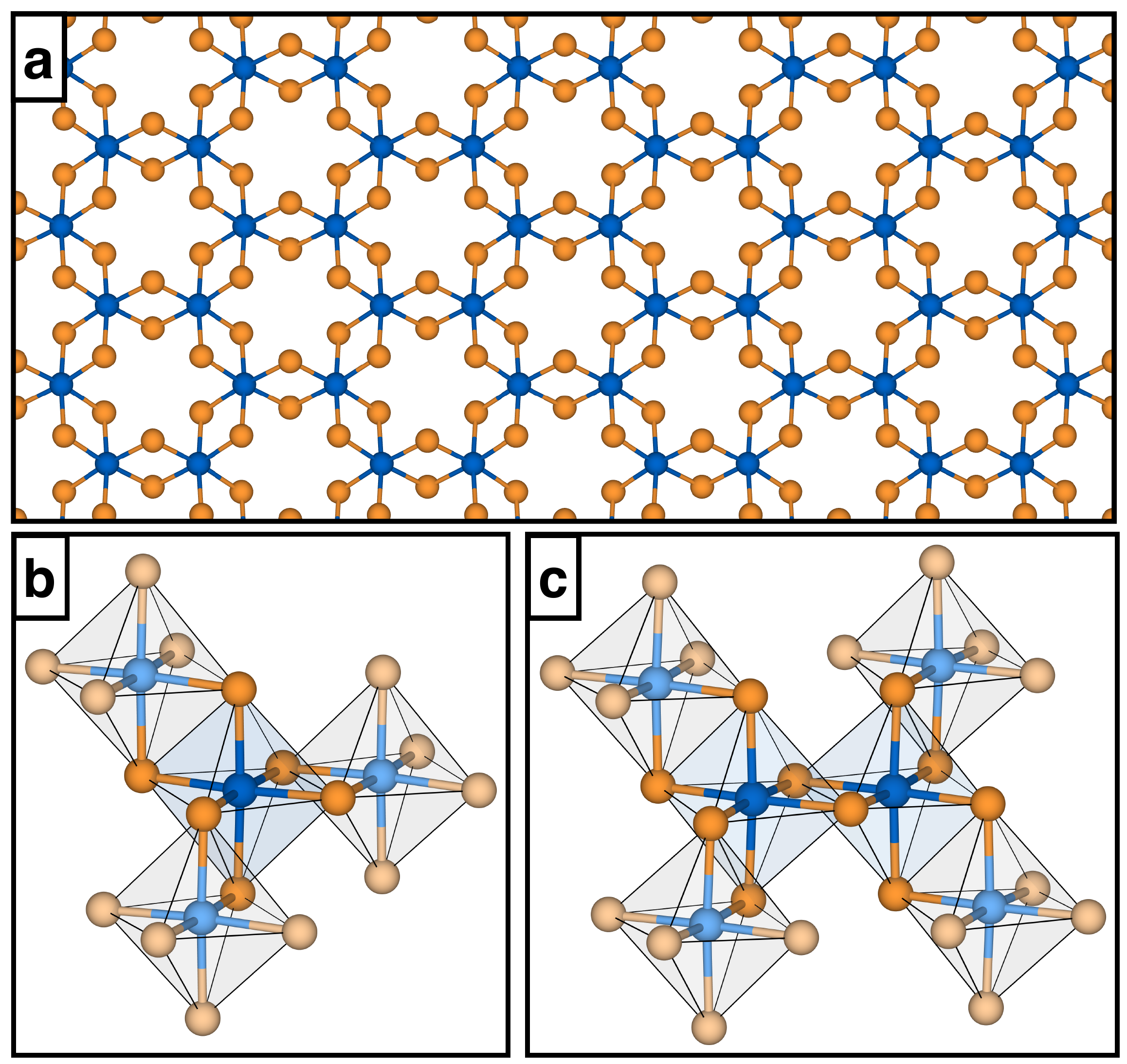}
    \caption{(a) Crystal structure of monolayer Cr$X_3$ ($X =$ Cl, Br, I). Blue and orange spheres represent chromium and halogen atoms, respectively. (b) Embedding model used in the determination of the multiplet structures and intra-site magnetic interactions. (c) Embedding model used in the determination of the inter-site magnetic interactions. In panels (b-c), dark (light) colors indicate the atoms that are treated with many-body (Hartree-Fock) wavefunctions. The models are embedded in an array of points charges (not shown) to reproduce the crystalline environment and ensure charge neutrality. \label{Fig1}}
\end{figure*}

We perform many-body wavefunction calculations on electrostatically embedded finite-size models constructed from the experimental bulk crystal structures of Cr$X_3$ ($X$ $=$ Cl, Br, I) materials \cite{McGuire2017, Klinkova1980, McGuire2015}. As shown in Figure \ref{Fig1}(a), the crystal structure of Cr$X_3$ magnets consists of layers in which  Cr atoms form a honeycomb network and are six-fold coordinated by halogen atoms. The finite-size models adopted in our calculations comprise a central unit hosting either one [Figure \ref{Fig1}(b) and Supporting Figure S1(a)] or two [Figure \ref{Fig1}(c)] octahedra treated with multireference or multiconfigurational wavefunctions, surrounded by the nearest-neighbor octahedra. These latter octahedra account for the charge distribution in the vicinity of the central unit and are treated at the Hartree-Fock level. The crystalline environment is restored by embedding each model in an array of point charges fitted to recreate the long-range Madelung electrostatic  potential \cite{Klintenberg2000}. Electron correlation effects in the central units are described at the complete-active-space self-consistent-field (CASSCF) and multi-reference configuration interaction (MRCI) levels of theory \cite{Helgaker2000}, including spin-orbit interactions. Technical details of our \emph{ab initio} calculations are discussed at length in Supporting Notes S1-S3. CASSCF approach is well suited to study systems involving degenerate or nearly degenerate configurations, where static correlation is important. In addition to static correlation, MRCI calculations also take into account dynamic correlation by considering single and double excitations from the $3d$ ($t_{2g}$) valence shells of Cr$^{3+}$ ions and the $p$ valence shells of the bridging halogen ligands. Therefore results at the MRCI level of theory provide a higher level of accuracy and should be used when comparing with the experimental results. Thus, we present both CASSCF and MRCI results to assess the relative importance of the two types of correlation effects. We remark that embedded finite-size model approaches have proven effective to describe electronic excitations and magnetic interactions in a rich variety of strongly correlated insulators \cite{Moreira2006, Malrieu2014}, including $d$-electron lattices \cite{Bogdanov2011, Katukuri2012, Moreira1999, Bogdanov2013}, owing to their localized nature.

\section{Results and Discussion} 
\textbf{Multiplet structures of Cr$^{3+}$ ions in Cr$X_3$.} We begin our work by establishing the Cr$^{3+}$ ground state and $d$-shell excitations in Cr$X_3$. In Table \ref{Table1}, we give the multiplet structure, as obtained at the CASSCF and MRCI levels using the one-site model shown in Figure 1(b). Additional details of our calculations are provided in the Supporting Note S1. According to the crystal-field theory, the octahedral environment encaging each Cr$^{3+}$ ion lifts the energy degeneracy of the $d$-orbital manifold, giving rise to a lower-lying triply degenerate $t_{2g}$ and a higher-lying doubly degenerate $e_g$ group of levels. The ground state of the halides is the high-spin singlet state $^4A_2$ dominated by the ($t^3_{2g}e^0_{g}$) configuration, in compliance with Hund's rule of maximum multiplicity. The half-filled $t_{2g}$ subshell renders each Cr$^{3+}$ ion a spin-3/2 center, in line with the magnetization saturation of 3 $\mu\textsubscript{B}$ observed in SQUID magnetometry measurements \cite{Huang2017}. The first excited state $^4T_2$ consists of dominant contributions from the configuration that comprises a pair of electrons in the $t_{2g}$ orbitals and a single electron in the $e_{g}$ orbitals. In the case of CrCl$_3$, this leads to a splitting between the $t_{2g}$ and $e_{g}$ orbitals as large as $\sim$1.6 eV at CASSCF level, which slightly increases to $\sim$1.7 eV at the MRCI level. Consistently with the order of the ligand strength subsumed in the spectrochemical series, such a splitting decreases in energy as the size of the ligand increases (that is, moving from CrCl$_3$ to CrI$_3$). In both CrBr$_3$ and CrI$_3$ the calculated splitting is $\sim$1.5~eV at the MRCI level of theory. The reason for this traces back to the more extended electron density in the heavier halogen atom and the consequently longer transition metal-to-ligand distances. Indeed, the  $t_{2g}$-$e_g$ splitting is sensitive to structural deformations of the octahedral cage and increases (decreases) when the crystal is subjected to moderate tensile (compressive) in-plain lattice strain, as we show for the representative case of CrCl$_3$ in Supporting Table S1. 

We gain theoretical insight into the recent x-ray spectroscopy experiments \cite{Shao2021} probing electronic excitations in Cr$X_3$. To that end, we simulate the Cr $L_{3,2}$-edge x-ray absorption spectra (XAS) and $L_{3}$-edge resonant inelastic x-ray scattering (RIXS) spectra of CrCl$_{3}$ and CrI$_{3}$ using CASSCF calculations and the minimal finite-size model shown in Supporting Figure S1(a). The resulting $d$-shell excitation energies of the Cr$^{3+}$ ion are listed in Supporting Tables S2 and S3. We consider the same scattering geometry used in recent experiments by Shao \emph{et al.} in Ref.\ \cite{Shao2021} and given in Supporting Figure S1(b). Additional details of these calculations are given in Supporting Note S2. We remark that similar computational schemes have been validated against experimental measurements for diverse magnetic compounds \cite{Nikolary_d9_rixs, Li_2021, Fabbris_2017}.
The simulated RIXS and XAS spectra of CrCl$_3$ and CrI$_3$ are shown in Figure 2. The excellent agreement between the theoretical and experimental spectra allows us to assign the electronic excitations observed. In Figures~2(a-b), the $L$-edge absorption process involves a Cr $2p$ electron that is promoted to the empty $3d$ orbitals ($2p \rightarrow 3p$), leading to the final  $2p^53d^4$ configuration ($2p^63d^3 \rightarrow 2p^53d^4$). This transition results in the two main features in the XAS spectra located at incident photon energies of  574$-$580~eV and 582$-$588~eV, which correspond to the Cr $L_3$-  and $L_2$-edges, respectively. These two features are separated by $\sim$9~eV due to spin-orbit interactions in the presence of the $2p$ core-hole in the XAS final states \cite{Laan_1992}. On the other hand, spin-orbit interactions weakly affect the Cr $3d^3$ valence states; see Supporting Tables S2 and S3.  

Due to the core-hole broadening in $L$-edge XAS, the onsite $d$-$d$ excitations are not well-resolved or determined by comparing our calculations with the experimental XAS spectra.
We further simulate the $L_3$-edge RIXS spectra to resolve the otherwise elusive $d$-shell excitations experimentally observed. This is accomplished by setting the incident energy, $E\textsubscript{in}$, at the maximum intensity of $L_3$-edge XAS, i.e., $E\textsubscript{in}$ = 576.6~eV for CrCl$_3$ and $E\textsubscript{in}$= 576.0~eV for CrI$_3$, and determining the RIXS intensities of $\pi$-polarization at incident angle $50^{\circ}$, similarly to experiments. As we restrict ourselves to on-site $d$-shell excitations and neglect ligand-metal charge transfer states, we focus on energy losses lower than 4~eV. The simulated RIXS spectra are shown in Figure~2(c$-$d) and, in analogy with experiments, exhibit three main features.  For CrCl$_3$, the first excitation peak at $\sim$1.6 eV corresponds to the $t_{2g} - e_g$ transitions associated with spin quartet states $^4T_2$($t^2_{2g}e^1_g$). 
The second excitation peak occurring in the 2.2$-$2.7 eV energy range is dominated by the $^4T_1$ ($t^2_{2g}e^1_g$) states, with a slight spectral weight contribution from the $^2E$ ($t^3_{2g}e^0_g$) and $^2T_1$ ($t^3_{2g}e^0_g$) excited states at 2.3$-$2.4~eV. 
The feature arising at $\sim$3.2~eV is related to the $^2T_{2}$ ($t^3_{2g}$) states.
The remaining higher energy loss feature from 3.5 $\sim$ 4.0 eV is attributed to
the $t_{2g} - e_g$ transitions associated with spin doublet states ($^2A$, $^2T_1$, and $^2T_2$).
Similarly, for CrI$_3$ the first excitation peak at $\sim$1.5~eV is assigned to the $t_{2g} - e_g$ transitions associated with spin quartet states $^4T_2$ ($t^2_{2g}e^1_g$). As compared to CrCl$_3$, however, the first excitation peak shift to lower energy loss  by $\sim$0.2~eV, both in the simulated and experimental spectra. The second feature at $\sim$2.3 eV is contributed by the $^4T_1$ ($t^2_{2g}e^1_g$) and $^2T_1$ ($t^3_{2g}$) states. At energy losses larger than 2.5~eV, the feature becomes broad and the $t_{2g} - e_g$ transitions
associated with spin doublet states are dominated.

\medskip
\textbf{Magnetic interactions in Cr$X_3$.} Next, we focus on magnetic interactions in Cr$X_3$ materials. We write the effective spin Hamiltonian $\mathcal{H}$ as

\begin{equation}
\mathcal{H} = \mathcal{H}^1 + \mathcal{H}^2,
\end{equation}
where $\mathcal{H}^1$ encodes the contribution of the interactions within sites of spin $\vec{S}_i$, while $\mathcal{H}^2$ describes the coupling between the $i$-th and $j$-th nearest-neighbor sites bearing spins $\vec{S}_i$ and $\vec{S}_j$, respectively. The intra-site contribution $\mathcal{H}^1$ takes the form

\begin{equation}
\mathcal{H}^1 =  \sum_{i} \vec{S_i} \cdot\bar{\bar{D}}_{\rm sia}\cdot \vec{S_i}   +  \sum_i \mu\textsubscript{B} \vec{B} \cdot \bar{\bar{g}} \cdot \vec{S_i},
\end{equation}
where $\bar{\bar{D}}_{\rm sia}$ is the single-ion anisotropy tensor and  $\bar{\bar{g}}$ is the $g$-tensor in the Zeeman term that emerges upon the application of an external magnetic field $\vec{B}$. The inter-site contribution $\mathcal{H}^2$ reads

\begin{equation}
\mathcal{H}^2 =  J{_1}  \sum_{i<j}(\vec{S_i} \cdot \vec{S_j}) +   J{_2}  \sum_{i<j}(\vec{S_i} \cdot \vec{S_j})^2 + \sum_{i<j} \vec{S_i} \cdot \bar{\bar{\Gamma}} \cdot \vec{S_j},
\end{equation}
with $J_1$ and $J_2$ being the bilinear and biquadratic isotropic exchange couplings, respectively, and $\bar{\bar{ \Gamma}}$ the symmetric anisotropic tensor. We remark that the nearest-neighbor Dzyaloshinskii-Moriya interaction vanishes due to the centrosymmetric crystal structure of Cr$X_3$ \cite{Moriya1960}. Assuming a local Kitaev frame where the $z$-axis is perpendicular to the Cr$_2$$X_2$ plaquette for each Cr$-$Cr bond \cite{Yadav_2018, Yadav_2016}, $\bar{\bar{\Gamma}}$ can be expressed as
 \begin{equation}
 \bar{\bar{\Gamma}} = 
\begin{pmatrix} 
0 & \Gamma_{xy} & -\Gamma_{yz} \\ 
\Gamma_{xy}  & 0 & \Gamma_{yz}  \\ 
-\Gamma_{yz} & \Gamma_{yz} & K
\end{pmatrix},
 \end{equation}
where $K$ is the Kitaev interaction parameter. 

We determine the intra-site interactions appearing in $\mathcal{H}^1$, {\it i.e.} the single-ion anisotropy and the $g$-tensor, relying on the one-site model shown in Figure 1(b). Further details of our calculations are given in Supporting Note S1. Our results are presented in Table \ref{Table2}. The single-ion anisotropy quantifies the zero-field splitting of the ground state that stems from spin-orbit and crystal-field effects. To assess this quantity, we rely on the methodology developed in Ref.\ \cite{Maurice2009} taking advantage of the multiplet structures listed in Table \ref{Table1} and corresponding wavefunctions.  In brief, the mixing of the low-lying $^4A_2$ states with the higher-lying states is treated perturbatively and the spin-orbit wavefunctions related to the high-spin $t^3_{2g}$ configuration are projected onto the space spanned by the $^4A_2$ $|S,M_s\rangle$ states. The orthonormalized projections of the low-lying quartet wavefunctions $\tilde{\psi}_k$ and the corresponding eigenvalues $E_k$ are used to construct the effective Hamiltonian $\tilde{H}_{\rm eff}=\Sigma_k E_k |\tilde{\psi}_k\rangle \langle \tilde{\psi}_k|$. A one-to-one correspondence between $\tilde{H}_{\rm eff}$ and the model Hamiltonian $\tilde{H}_{\rm mod}=\vec{S_i} \cdot\bar{\bar{D}}_{\rm sia}\cdot \vec{S_i}$ leads to the ${ \bar{\bar{D}}_{\rm sia}}$ tensor (listed in Supporting Table S4), which is then diagonalized and the axial parameter $A_{\rm sia}$ obtained as $A_{\rm sia}=D_{\rm sia}^{zz}-(D_{\rm sia}^{xx}+D_{\rm sia}^{yy})/2$. At the MRCI level of theory, we find $A_{\rm sia} = -0.03$ meV for CrCl$_3$, $ -0.08$ meV for CrBr$_3$ and $ -0.12$ meV for CrI$_3$. The negative sign of the axial parameter corresponds an easy axis of magnetization. In all the three cases we find an easy axis pointing perpendicular to the honeycomb plane.
The considerably large magnitude of the single-ion anisotropy in CrI$_3$ compared to CrCl$_3$ reflects the strong spin-orbit interactions inherent to the heavier ligands \cite{Lado2017}.


Another important contribution to the anisotropy originates from magnetic dipole-dipole interactions that are not accounted for in the electronic structure calculations \cite{akhiezer1968spin,aharoni2000introduction}. This contribution, however, should not be neglected for monolayer as well as layered bulk magnetic materials 
as a result of magnetic shape anisotropy. The magnetic dipolar anisotropy for any given lattice can be determined as:

\begin{equation}
\mathcal{H}_{\rm dip} =  (g_e\mu_{\mathrm B})^2 \sum_{i < j} \left[\frac{(|\vec{r}_{ij}|^2(\vec{S}_i \cdot \vec{S}_j)- 3 (\vec{S}_i \cdot \vec{r}_{ij})(\vec{S}_j \cdot \vec{r}_{ij})}{|\vec{r}_{ij}|^5}\right],
\end{equation}
where $\vec{S}_{i(j)}$ is spin at site $i (j)$ and $\vec{r_{ij}}$ is the vector connecting sites $i$ and $j$. This expression can be rewritten in a form that is similar to Equation 2:

\begin{equation}
\mathcal{H}_{\rm dip} =  \sum_{i < j} \vec{S_i} \cdot \bar{\bar{D}}_{\rm dip} \cdot \vec{S_j} ,
\end{equation}
where $\bar{\bar{D}}_{\rm dip}$ is a tensor. The matrix elements $D_{\rm dip}^{\alpha\beta}$ ($\alpha,\beta = x, y, z$) are computed as

\begin{equation}
D_{\rm dip}^{\alpha\beta} = (g_e\mu_{\mathrm B})^2 \sum_{i<j} \left[\frac{|\vec{r}_{ij}|^2(\hat{e}^\alpha_i \cdot \hat{e}^\beta_j)- 3 (\hat{e}^\alpha_i \cdot \vec{r}_{ij})( \hat{e}^\beta_j \cdot \vec{r}_{ij})}{|\vec{r}_{ij}|^5}\right],
\end{equation}
with $\hat{e}^\alpha_i$ being the unit vector at site $i$ pointing along $\alpha = x, y, z$.
In order to avoid summing over an infinite number of pairs of spins we used the approach described in Ref.\,\cite{Kim2021}, which is based on performing summation upto a certain cutoff distance between the spins and carefully evaluating the convergence of the matrix elements of $\bar{\bar{D}}_{\rm dip}$ with respect to this cutoff. Supporting Figure S2 presents the converged values of $D_{\rm dip}^{\alpha\alpha}$ for the $x$, $y$, $z$ orientation of magnetic moments in CrCl$_3$, CrBr$_3$ and CrI$_3$. Similar to the single-ion anisotropy axial parameter, we obtain the axial parameter for the magnetic dipolar anisotropy $A_{\rm dip}=D_{\rm dip}^{zz}-(D_{\rm dip}^{xx}+D_{\rm dip}^{yy})/2$. The $A_{\rm dip}$ parameters for all three chromium trihalides are listed in Table \ref{shape_aniso}.
The magnitudes of $A_{\rm dip}$ decrease along the series as a result of a slight increase of the in-plane lattice constants and are larger for monolayers as compared to bulk materials.
Most importantly, these values are positive, which is consistent with the expectation of the demagnetization energy being lower for the in-plane magnetization in slab geometries, and are of magnitude comparable to that of single-ion anisotropy.
This allows us to conclude that the overall magnetic anisotropy in chromium trihalides is dictated by the interplay between these two competing contributions.
In the heavier CrI$_3$ single-ion anisotropy dominates ($A_{\rm sia}+A_{\rm dip} = -67$~$\mu$eV) establishing as the (out-of-plane) easy-axis ferromagnet, while the opposite is true for CrCl$_3$ ($A_{\rm sia}+A_{\rm dip} = 54$~$\mu$eV) that we conclude to be an easy-plane ferromagnet. The case of CrBr$_3$ is borderline with the marginally larger contribution from single-ion anisotropy, which establishes this material as an easy-axis magnet with weak anisotropy.  Our findings are in full agreement with recent experiments\,\cite{Huang2017, McGuire2017,Cai2019,Kim2019}. 


There is, however, one subtle point \cite{maleev1976}. The transition from Equation 5 to 6 is rigorous only for homogeneous distribution of magnetization plus ellipsoidal shape of the sample \cite{aharoni2000introduction}. The spin-wave spectrum for ferromagnets with an easy-plane anisotropy and for those with dipole-dipole interactions are essentially different, and for the latter the spin-wave energy tends to zero at zero wave vector much slower \cite{akhiezer1968spin,maleev1976}. Also, due to the long-range character of dipole-dipole interactions the formal conditions of applicability of Mermin-Wagner-Hohenberg theorem are not fulfilled. Maleev suggested \cite{maleev1976} that in 2D ferromagnets with dipole-dipole interactions one has the true long-range order at low enough temperatures, contrary to the ``almost broken symmetry'' of Berezinskii-Kosterlitz-Thouless type \cite{kleinert_book,PhysRevB.60.2990} for the easy-plane ferromagnets. Later this suggestion was confirmed within the framework of self-consistent spin-wave theory \cite{PhysRevB.71.024427}. This means that for all three Cr$X_3$ ($X =$ Cl, Br, I) ferromagnets one can expect the true long-range ordered state below the Curie temperature. 

Of crucial importance to understand the response of  Cr$X_3$ to external magnetic fields is the $g$-tensor. We determine this quantity following the scheme devised in Ref.\ \cite{Bolvin2006}. Many-body wavefunction calculations enable to extract the matrix elements of the total magnetic moment operator ${\bf \hat{\mu}}$ in the basis of the spin-orbit coupled eigenstates ${\bf \hat{\mu}}=-\mu_B(g_e\hat{{S}}+ \hat{{L}})$, where $\hat{{L}}$ is the angular momentum operator, $\hat{{S}}$  the spin operator, and $g_e$ the free-electron Land\'e factor for a given magnetic site.  The Zeeman Hamiltonian in terms of total magnetic moment, $H\textsubscript{Z}=-\hat{{\mu}} \cdot \vec{B}$, is then mapped onto the Zeeman contribution to $\mathcal{H}^1$ given in Equation 2. To obtain the $g$-factors, we use an active space that includes all five $d$ orbitals of the Cr$^{3+}$ ion averaged over five quartets and seven doublets.  The $g$-factors at both CASSCF and MRCI level of theory are listed in Table \ref{Table2}. These values match those determined in recent electron spin resonance experiments \cite{Singamaneni2020}, that is 1.49 and 1.96 for CrCl$_3$ and CrI$_3$, respectively. The slight anisotropy ($\sim$1\%) of the calculated $g$-tensors arises from the mild trigonal distortion of the octahedral cage occurring in Cr$X_3$ ferromagnets.

Finally, we determine the inter-site interactions appearing in the spin Hamiltonian $\mathcal{H}^2$, i.e., the bilinear and biquadratic isotropic exchange couplings $J_1$ and $J_2$, respectively,  together with the symmetric anisotropic tensor $\bar{\bar{{\Gamma}}}$. We rely on the two-site model shown in Figure 1(c) and give further technicalities of our calculations in Supporting Note S3. We map the resulting \emph{ab initio} Hamiltonian onto the anisotropic biquadratic model Hamiltonian given in Equation 3, which involves sixteen spin-orbit states corresponding to one septet, one quintet, one triplet, and a singlet. The mapping is accomplished according to the procedure described in Ref.\ \cite{Bogdanov2013}. The resulting magnetic interactions are listed in Table \ref{Table3}. Additional results obtained on a simpler isotropic bilinear Hamiltonian are provided in Supporting Table S5.

The dominant term among the inter-site interactions is the bilinear isotropic exchange coupling $J_1$, the negative sign of which indicates a parallel spin interaction between the spin-3/2 centers. This is fully consistent with the observations of an intra-layer ferromagnetic order probed by magneto-optical Kerr effect in CrI$_3$ \cite{Huang2017} and temperature-dependence transport measurements in CrCl$_3$ \cite{Cai2019}. The strength of $J_1$ is largely dependent on whether the CASSCF or MRCI level of theory is adopted due to the different types of exchange mechanisms involved in these approaches. At the CASSCF level, the main mechanism is the direct exchange between the transition metal atoms. As a result, $J_1$ is slightly larger in CrCl$_3$ and CrBr$_3$ as compared to CrI$_3$ as a consequence of the shorter distance between the Cr$^{3+}$ ions in the former trihalides. At the MRCI level, however, the super-exchange pathway through the non-magnetic ligands becomes operative, significantly strengthening $J_1$ which attains the value of $-0.97$~meV in CrCl$_3$, $-1.21$~meV in CrBr$_3$ and $-1.38$~meV in CrI$_3$. This increase in the magnitude of $J_1$ quantifies the super-exchange contribution to the bilinear exchange interaction, which is greater in CrI$_3$ (0.78~meV) as compared to CrCl$_3$ (0.33~meV) owing to the more extended $p$ valence orbitals of the iodine atoms compared to the chlorine atoms. Hence, the sign of the bilinear exchange interaction can be qualitatively rationalized in terms of the rule devised by Goodenough \cite{Goodenough1958} and Kanamori \cite{Kanamori1959}, which points to a ferromagnetic coupling when the Cr$-$$X$$-$Cr bond angle approaches $90$\textsuperscript{o}. The calculated values of the bilinear exchange interaction obtained at the MRCI level are in excellent agreement with certain recent experimental estimates, e.g., $J_1 = -0.92$~meV for CrCl$_3$ \cite{Kim2019} and  $-1.42$~meV for CrI$_3$ \cite{Cenker2021}, and their trend is consistent with the significantly lower critical temperature observed in the former magnet ($\sim$16 K \cite{Kim2019}) than the latter (45 K \cite{Huang2017}).

Importantly, we reveal a relatively sizable biquadratic exchange couplings $J_2$, thus indicating that ferromagnetic two-electron hopping processes are active in chromium trihalides \cite{Kartsev2020, Wahab2021, Ni2021}. The strength of $J_2$ appears to be weakly dependent on the nature of the halogen atom and on whether or not super-exchange effects are described. This possibly suggests that, contrary to bilinear interactions, multispin interactions occur via the direct exchange between the Cr$^{3+}$ ions. Both the isotropic bilinear and biquadratic exchange coupling largely exceed the anisotropic exchange interactions. Among the anisotropic exchange couplings $\bar{\bar{\Gamma}}$, only the Kitaev interaction in CrI$_3$ is non-vanishing ($-0.05$ meV), with all the off-diagonal terms being negligible. Irrespective of the nature of the halogen ligands, however, our results indicate a very small Kitaev-to-Heisenberg ratio, lending support to earlier findings on CrI$_3$ \cite{PhysRevB.102.115162, Lado2017, Torelli_2018}, and casting doubts on others \cite{Xu2018, Lee2020}. 

\section{Conclusions} 
Using multiconfigurational and multireference wavefunctions, we have achieved a description of reference accuracy for the electronic excitations and the magnetic Hamiltonian parameters  in chromium trihalides, as demonstrated by the excellent agreement with a variety of spectroscopy experiments. Our work serves as a reference for future computational studies of Cr$X_3$ and establishes many-body wavefunction based calculations on embedded clusters as an effective approach to elucidating the nature of electronic and spin interactions in two-dimensional semiconducting magnets. Our calculations show that the magnetization direction which is determined by the interplay of out-of-plane magnetocrystalline anisotropy and magnetic dipole-dipole interaction favoring in-plane magnetization is clearly out-of-plane for CrI$_3$ and in-plane for CrCl$_3$ whereas for CrBr$_3$ these two contributions almost exactly compensate one another, with a small out-of-plane resulting anisotropy. Note however that since out-of-plane direction in CrCl$_3$ is provided by long-range dipole-dipole interactions Mermin-Wagner-Hohenberg theorem is not applicable in this case and one case expect true long-range order at low temperatures \cite{maleev1976,PhysRevB.71.024427}.  

\section{Acknowledgments}
R.Y. is supported by the Swiss National Science Foundation - Sinergia Network NanoSkyrmionics (Grant No.\ CRSII5-171003) and NCCR MARVEL. M.\ P.\ is  supported by the Swiss National Science Foundation (SNSF) through the Early Postdoc.Mobility program (Grant No.\ P2ELP2-191706). The work of M.I.K. was supported by the ERC Synergy Grant, Project No. 854843 FASTCORR. L. X. thanks U. Nitzsche for technical assistance.

 \bibliography{ms}

\begin{thebibliography}{69}%
\makeatletter
\providecommand \@ifxundefined [1]{%
 \@ifx{#1\undefined}
}%
\providecommand \@ifnum [1]{%
 \ifnum #1\expandafter \@firstoftwo
 \else \expandafter \@secondoftwo
 \fi
}%
\providecommand \@ifx [1]{%
 \ifx #1\expandafter \@firstoftwo
 \else \expandafter \@secondoftwo
 \fi
}%
\providecommand \natexlab [1]{#1}%
\providecommand \enquote  [1]{``#1''}%
\providecommand \bibnamefont  [1]{#1}%
\providecommand \bibfnamefont [1]{#1}%
\providecommand \citenamefont [1]{#1}%
\providecommand \href@noop [0]{\@secondoftwo}%
\providecommand \href [0]{\begingroup \@sanitize@url \@href}%
\providecommand \@href[1]{\@@startlink{#1}\@@href}%
\providecommand \@@href[1]{\endgroup#1\@@endlink}%
\providecommand \@sanitize@url [0]{\catcode `\\12\catcode `\$12\catcode
  `\&12\catcode `\#12\catcode `\^12\catcode `\_12\catcode `\%12\relax}%
\providecommand \@@startlink[1]{}%
\providecommand \@@endlink[0]{}%
\providecommand \url  [0]{\begingroup\@sanitize@url \@url }%
\providecommand \@url [1]{\endgroup\@href {#1}{\urlprefix }}%
\providecommand \urlprefix  [0]{URL }%
\providecommand \Eprint [0]{\href }%
\providecommand \doibase [0]{http://dx.doi.org/}%
\providecommand \selectlanguage [0]{\@gobble}%
\providecommand \bibinfo  [0]{\@secondoftwo}%
\providecommand \bibfield  [0]{\@secondoftwo}%
\providecommand \translation [1]{[#1]}%
\providecommand \BibitemOpen [0]{}%
\providecommand \bibitemStop [0]{}%
\providecommand \bibitemNoStop [0]{.\EOS\space}%
\providecommand \EOS [0]{\spacefactor3000\relax}%
\providecommand \BibitemShut  [1]{\csname bibitem#1\endcsname}%
\let\auto@bib@innerbib\@empty
\bibitem [{\citenamefont {Burch}\ \emph {et~al.}(2018)\citenamefont {Burch},
  \citenamefont {Mandrus},\ and\ \citenamefont {Park}}]{Burch2018}%
  \BibitemOpen
  \bibfield  {author} {\bibinfo {author} {\bibfnamefont {K.~S.}\ \bibnamefont
  {Burch}}, \bibinfo {author} {\bibfnamefont {D.}~\bibnamefont {Mandrus}}, \
  and\ \bibinfo {author} {\bibfnamefont {J.-G.}\ \bibnamefont {Park}},\ }\href
  {\doibase 10.1038/s41586-018-0631-z} {\bibfield  {journal} {\bibinfo
  {journal} {Nature}\ }\textbf {\bibinfo {volume} {563}},\ \bibinfo {pages}
  {47} (\bibinfo {year} {2018})}\BibitemShut {NoStop}%
\bibitem [{\citenamefont {Gibertini}\ \emph {et~al.}(2019)\citenamefont
  {Gibertini}, \citenamefont {Koperski}, \citenamefont {Morpurgo},\ and\
  \citenamefont {Novoselov}}]{Gibertini2019}%
  \BibitemOpen
  \bibfield  {author} {\bibinfo {author} {\bibfnamefont {M.}~\bibnamefont
  {Gibertini}}, \bibinfo {author} {\bibfnamefont {M.}~\bibnamefont {Koperski}},
  \bibinfo {author} {\bibfnamefont {A.~F.}\ \bibnamefont {Morpurgo}}, \ and\
  \bibinfo {author} {\bibfnamefont {K.~S.}\ \bibnamefont {Novoselov}},\ }\href
  {\doibase 10.1038/s41565-019-0438-6} {\bibfield  {journal} {\bibinfo
  {journal} {Nature Nanotechnology}\ }\textbf {\bibinfo {volume} {14}},\
  \bibinfo {pages} {408} (\bibinfo {year} {2019})}\BibitemShut {NoStop}%
\bibitem [{\citenamefont {Gong}\ and\ \citenamefont {Zhang}(2019)}]{Gong2019}%
  \BibitemOpen
  \bibfield  {author} {\bibinfo {author} {\bibfnamefont {C.}~\bibnamefont
  {Gong}}\ and\ \bibinfo {author} {\bibfnamefont {X.}~\bibnamefont {Zhang}},\
  }\href {\doibase 10.1126/science.aav4450} {\bibfield  {journal} {\bibinfo
  {journal} {Science}\ }\textbf {\bibinfo {volume} {363}},\ \bibinfo {pages}
  {706} (\bibinfo {year} {2019})}\BibitemShut {NoStop}%
\bibitem [{\citenamefont {Mak}\ \emph {et~al.}(2019)\citenamefont {Mak},
  \citenamefont {Shan},\ and\ \citenamefont {Ralph}}]{Mak2019}%
  \BibitemOpen
  \bibfield  {author} {\bibinfo {author} {\bibfnamefont {K.~F.}\ \bibnamefont
  {Mak}}, \bibinfo {author} {\bibfnamefont {J.}~\bibnamefont {Shan}}, \ and\
  \bibinfo {author} {\bibfnamefont {D.~C.}\ \bibnamefont {Ralph}},\ }\href@noop
  {} {\bibfield  {journal} {\bibinfo  {journal} {Nature Reviews Physics}\
  }\textbf {\bibinfo {volume} {1}},\ \bibinfo {pages} {646} (\bibinfo {year}
  {2019})}\BibitemShut {NoStop}%
\bibitem [{\citenamefont {Mermin}\ and\ \citenamefont
  {Wagner}(1966)}]{Mermin1966}%
  \BibitemOpen
  \bibfield  {author} {\bibinfo {author} {\bibfnamefont {N.~D.}\ \bibnamefont
  {Mermin}}\ and\ \bibinfo {author} {\bibfnamefont {H.}~\bibnamefont
  {Wagner}},\ }\href {\doibase 10.1103/PhysRevLett.17.1133} {\bibfield
  {journal} {\bibinfo  {journal} {Physical Review Letters}\ }\textbf {\bibinfo
  {volume} {17}},\ \bibinfo {pages} {1133} (\bibinfo {year}
  {1966})}\BibitemShut {NoStop}%
\bibitem [{\citenamefont {Hohenberg}(1967)}]{Hohenberg1967}%
  \BibitemOpen
  \bibfield  {author} {\bibinfo {author} {\bibfnamefont {P.~C.}\ \bibnamefont
  {Hohenberg}},\ }\href {\doibase 10.1103/PhysRev.158.383} {\bibfield
  {journal} {\bibinfo  {journal} {Physical Review}\ }\textbf {\bibinfo {volume}
  {158}},\ \bibinfo {pages} {383} (\bibinfo {year} {1967})}\BibitemShut
  {NoStop}%
\bibitem [{\citenamefont {Soriano}\ \emph {et~al.}(2020)\citenamefont
  {Soriano}, \citenamefont {Katsnelson},\ and\ \citenamefont
  {Fern{\'{a}}ndez-Rossier}}]{Soriano2020}%
  \BibitemOpen
  \bibfield  {author} {\bibinfo {author} {\bibfnamefont {D.}~\bibnamefont
  {Soriano}}, \bibinfo {author} {\bibfnamefont {M.~I.}\ \bibnamefont
  {Katsnelson}}, \ and\ \bibinfo {author} {\bibfnamefont {J.}~\bibnamefont
  {Fern{\'{a}}ndez-Rossier}},\ }\href {\doibase 10.1021/acs.nanolett.0c02381}
  {\bibfield  {journal} {\bibinfo  {journal} {Nano Letters}\ }\textbf {\bibinfo
  {volume} {20}},\ \bibinfo {pages} {6225} (\bibinfo {year}
  {2020})}\BibitemShut {NoStop}%
\bibitem [{\citenamefont {Huang}\ \emph {et~al.}(2017)\citenamefont {Huang},
  \citenamefont {Clark}, \citenamefont {Navarro-Moratalla}, \citenamefont
  {Klein}, \citenamefont {Cheng}, \citenamefont {Seyler}, \citenamefont
  {Zhong}, \citenamefont {Schmidgall}, \citenamefont {McGuire}, \citenamefont
  {Cobden}, \citenamefont {Yao}, \citenamefont {Xiao}, \citenamefont
  {Jarillo-Herrero},\ and\ \citenamefont {Xu}}]{Huang2017}%
  \BibitemOpen
  \bibfield  {author} {\bibinfo {author} {\bibfnamefont {B.}~\bibnamefont
  {Huang}}, \bibinfo {author} {\bibfnamefont {G.}~\bibnamefont {Clark}},
  \bibinfo {author} {\bibfnamefont {E.}~\bibnamefont {Navarro-Moratalla}},
  \bibinfo {author} {\bibfnamefont {D.~R.}\ \bibnamefont {Klein}}, \bibinfo
  {author} {\bibfnamefont {R.}~\bibnamefont {Cheng}}, \bibinfo {author}
  {\bibfnamefont {K.~L.}\ \bibnamefont {Seyler}}, \bibinfo {author}
  {\bibfnamefont {D.}~\bibnamefont {Zhong}}, \bibinfo {author} {\bibfnamefont
  {E.}~\bibnamefont {Schmidgall}}, \bibinfo {author} {\bibfnamefont {M.~A.}\
  \bibnamefont {McGuire}}, \bibinfo {author} {\bibfnamefont {D.~H.}\
  \bibnamefont {Cobden}}, \bibinfo {author} {\bibfnamefont {W.}~\bibnamefont
  {Yao}}, \bibinfo {author} {\bibfnamefont {D.}~\bibnamefont {Xiao}}, \bibinfo
  {author} {\bibfnamefont {P.}~\bibnamefont {Jarillo-Herrero}}, \ and\ \bibinfo
  {author} {\bibfnamefont {X.}~\bibnamefont {Xu}},\ }\href@noop {} {\bibfield
  {journal} {\bibinfo  {journal} {Nature}\ }\textbf {\bibinfo {volume} {546}},\
  \bibinfo {pages} {270} (\bibinfo {year} {2017})}\BibitemShut {NoStop}%
\bibitem [{\citenamefont {Sivadas}\ \emph {et~al.}(2018)\citenamefont
  {Sivadas}, \citenamefont {Okamoto}, \citenamefont {Xu}, \citenamefont
  {Fennie},\ and\ \citenamefont {Xiao}}]{Sivadas2018}%
  \BibitemOpen
  \bibfield  {author} {\bibinfo {author} {\bibfnamefont {N.}~\bibnamefont
  {Sivadas}}, \bibinfo {author} {\bibfnamefont {S.}~\bibnamefont {Okamoto}},
  \bibinfo {author} {\bibfnamefont {X.}~\bibnamefont {Xu}}, \bibinfo {author}
  {\bibfnamefont {C.~J.}\ \bibnamefont {Fennie}}, \ and\ \bibinfo {author}
  {\bibfnamefont {D.}~\bibnamefont {Xiao}},\ }\href {\doibase
  10.1021/acs.nanolett.8b03321} {\bibfield  {journal} {\bibinfo  {journal}
  {Nano Letters}\ }\textbf {\bibinfo {volume} {18}},\ \bibinfo {pages} {7658}
  (\bibinfo {year} {2018})}\BibitemShut {NoStop}%
\bibitem [{\citenamefont {Song}\ \emph {et~al.}(2019)\citenamefont {Song},
  \citenamefont {Fei}, \citenamefont {Yankowitz}, \citenamefont {Lin},
  \citenamefont {Jiang}, \citenamefont {Hwangbo}, \citenamefont {Zhang},
  \citenamefont {Sun}, \citenamefont {Taniguchi}, \citenamefont {Watanabe},
  \citenamefont {McGuire}, \citenamefont {Graf}, \citenamefont {Cao},
  \citenamefont {Chu}, \citenamefont {Cobden}, \citenamefont {Dean},
  \citenamefont {Xiao},\ and\ \citenamefont {Xu}}]{Song2019}%
  \BibitemOpen
  \bibfield  {author} {\bibinfo {author} {\bibfnamefont {T.}~\bibnamefont
  {Song}}, \bibinfo {author} {\bibfnamefont {Z.}~\bibnamefont {Fei}}, \bibinfo
  {author} {\bibfnamefont {M.}~\bibnamefont {Yankowitz}}, \bibinfo {author}
  {\bibfnamefont {Z.}~\bibnamefont {Lin}}, \bibinfo {author} {\bibfnamefont
  {Q.}~\bibnamefont {Jiang}}, \bibinfo {author} {\bibfnamefont
  {K.}~\bibnamefont {Hwangbo}}, \bibinfo {author} {\bibfnamefont
  {Q.}~\bibnamefont {Zhang}}, \bibinfo {author} {\bibfnamefont
  {B.}~\bibnamefont {Sun}}, \bibinfo {author} {\bibfnamefont {T.}~\bibnamefont
  {Taniguchi}}, \bibinfo {author} {\bibfnamefont {K.}~\bibnamefont {Watanabe}},
  \bibinfo {author} {\bibfnamefont {M.~A.}\ \bibnamefont {McGuire}}, \bibinfo
  {author} {\bibfnamefont {D.}~\bibnamefont {Graf}}, \bibinfo {author}
  {\bibfnamefont {T.}~\bibnamefont {Cao}}, \bibinfo {author} {\bibfnamefont
  {J.-H.}\ \bibnamefont {Chu}}, \bibinfo {author} {\bibfnamefont {D.~H.}\
  \bibnamefont {Cobden}}, \bibinfo {author} {\bibfnamefont {C.~R.}\
  \bibnamefont {Dean}}, \bibinfo {author} {\bibfnamefont {D.}~\bibnamefont
  {Xiao}}, \ and\ \bibinfo {author} {\bibfnamefont {X.}~\bibnamefont {Xu}},\
  }\href@noop {} {\bibfield  {journal} {\bibinfo  {journal} {Nature Materials}\
  }\textbf {\bibinfo {volume} {18}},\ \bibinfo {pages} {1298} (\bibinfo {year}
  {2019})}\BibitemShut {NoStop}%
\bibitem [{\citenamefont {Huang}\ \emph {et~al.}(2018)\citenamefont {Huang},
  \citenamefont {Clark}, \citenamefont {Klein}, \citenamefont {MacNeill},
  \citenamefont {Navarro-Moratalla}, \citenamefont {Seyler}, \citenamefont
  {Wilson}, \citenamefont {McGuire}, \citenamefont {Cobden}, \citenamefont
  {Xiao}, \citenamefont {Yao}, \citenamefont {Jarillo-Herrero},\ and\
  \citenamefont {Xu}}]{Huang2018}%
  \BibitemOpen
  \bibfield  {author} {\bibinfo {author} {\bibfnamefont {B.}~\bibnamefont
  {Huang}}, \bibinfo {author} {\bibfnamefont {G.}~\bibnamefont {Clark}},
  \bibinfo {author} {\bibfnamefont {D.~R.}\ \bibnamefont {Klein}}, \bibinfo
  {author} {\bibfnamefont {D.}~\bibnamefont {MacNeill}}, \bibinfo {author}
  {\bibfnamefont {E.}~\bibnamefont {Navarro-Moratalla}}, \bibinfo {author}
  {\bibfnamefont {K.~L.}\ \bibnamefont {Seyler}}, \bibinfo {author}
  {\bibfnamefont {N.}~\bibnamefont {Wilson}}, \bibinfo {author} {\bibfnamefont
  {M.~A.}\ \bibnamefont {McGuire}}, \bibinfo {author} {\bibfnamefont {D.~H.}\
  \bibnamefont {Cobden}}, \bibinfo {author} {\bibfnamefont {D.}~\bibnamefont
  {Xiao}}, \bibinfo {author} {\bibfnamefont {W.}~\bibnamefont {Yao}}, \bibinfo
  {author} {\bibfnamefont {P.}~\bibnamefont {Jarillo-Herrero}}, \ and\ \bibinfo
  {author} {\bibfnamefont {X.}~\bibnamefont {Xu}},\ }\href {\doibase
  10.1038/s41565-018-0121-3} {\bibfield  {journal} {\bibinfo  {journal} {Nature
  Nanotechnology}\ }\textbf {\bibinfo {volume} {13}},\ \bibinfo {pages} {544}
  (\bibinfo {year} {2018})}\BibitemShut {NoStop}%
\bibitem [{\citenamefont {Jiang}\ \emph {et~al.}(2018)\citenamefont {Jiang},
  \citenamefont {Li}, \citenamefont {Wang}, \citenamefont {Mak},\ and\
  \citenamefont {Shan}}]{Jiang2018}%
  \BibitemOpen
  \bibfield  {author} {\bibinfo {author} {\bibfnamefont {S.}~\bibnamefont
  {Jiang}}, \bibinfo {author} {\bibfnamefont {L.}~\bibnamefont {Li}}, \bibinfo
  {author} {\bibfnamefont {Z.}~\bibnamefont {Wang}}, \bibinfo {author}
  {\bibfnamefont {K.~F.}\ \bibnamefont {Mak}}, \ and\ \bibinfo {author}
  {\bibfnamefont {J.}~\bibnamefont {Shan}},\ }\href {\doibase
  10.1038/s41565-018-0135-x} {\bibfield  {journal} {\bibinfo  {journal} {Nature
  Nanotechnology}\ }\textbf {\bibinfo {volume} {13}},\ \bibinfo {pages} {549}
  (\bibinfo {year} {2018})}\BibitemShut {NoStop}%
\bibitem [{\citenamefont {Li}\ \emph {et~al.}(2019)\citenamefont {Li},
  \citenamefont {Jiang}, \citenamefont {Sivadas}, \citenamefont {Wang},
  \citenamefont {Xu}, \citenamefont {Weber}, \citenamefont {Goldberger},
  \citenamefont {Watanabe}, \citenamefont {Taniguchi}, \citenamefont {Fennie},
  \citenamefont {Mak},\ and\ \citenamefont {Shan}}]{Tingxin2019}%
  \BibitemOpen
  \bibfield  {author} {\bibinfo {author} {\bibfnamefont {T.}~\bibnamefont
  {Li}}, \bibinfo {author} {\bibfnamefont {S.}~\bibnamefont {Jiang}}, \bibinfo
  {author} {\bibfnamefont {N.}~\bibnamefont {Sivadas}}, \bibinfo {author}
  {\bibfnamefont {Z.}~\bibnamefont {Wang}}, \bibinfo {author} {\bibfnamefont
  {Y.}~\bibnamefont {Xu}}, \bibinfo {author} {\bibfnamefont {D.}~\bibnamefont
  {Weber}}, \bibinfo {author} {\bibfnamefont {J.~E.}\ \bibnamefont
  {Goldberger}}, \bibinfo {author} {\bibfnamefont {K.}~\bibnamefont
  {Watanabe}}, \bibinfo {author} {\bibfnamefont {T.}~\bibnamefont {Taniguchi}},
  \bibinfo {author} {\bibfnamefont {C.~J.}\ \bibnamefont {Fennie}}, \bibinfo
  {author} {\bibfnamefont {K.~F.}\ \bibnamefont {Mak}}, \ and\ \bibinfo
  {author} {\bibfnamefont {J.}~\bibnamefont {Shan}},\ }\href@noop {} {\bibfield
   {journal} {\bibinfo  {journal} {Nature Materials}\ }\textbf {\bibinfo
  {volume} {18}},\ \bibinfo {pages} {1303} (\bibinfo {year}
  {2019})}\BibitemShut {NoStop}%
\bibitem [{\citenamefont {Pizzochero}\ and\ \citenamefont
  {Yazyev}(2020)}]{Pizzochero2020a}%
  \BibitemOpen
  \bibfield  {author} {\bibinfo {author} {\bibfnamefont {M.}~\bibnamefont
  {Pizzochero}}\ and\ \bibinfo {author} {\bibfnamefont {O.~V.}\ \bibnamefont
  {Yazyev}},\ }\href {\doibase 10.1021/acs.jpcc.0c01873} {\bibfield  {journal}
  {\bibinfo  {journal} {The Journal of Physical Chemistry C}\ }\textbf
  {\bibinfo {volume} {124}},\ \bibinfo {pages} {7585} (\bibinfo {year}
  {2020})}\BibitemShut {NoStop}%
\bibitem [{\citenamefont {Webster}\ and\ \citenamefont
  {Yan}(2018)}]{PhysRevB.98.144411}%
  \BibitemOpen
  \bibfield  {author} {\bibinfo {author} {\bibfnamefont {L.}~\bibnamefont
  {Webster}}\ and\ \bibinfo {author} {\bibfnamefont {J.-A.}\ \bibnamefont
  {Yan}},\ }\href {\doibase 10.1103/PhysRevB.98.144411} {\bibfield  {journal}
  {\bibinfo  {journal} {Phys. Rev. B}\ }\textbf {\bibinfo {volume} {98}},\
  \bibinfo {pages} {144411} (\bibinfo {year} {2018})}\BibitemShut {NoStop}%
\bibitem [{\citenamefont {Wang}\ \emph {et~al.}(2020)\citenamefont {Wang},
  \citenamefont {Su}, \citenamefont {Yang}, \citenamefont {Zhang},\ and\
  \citenamefont {Zhang}}]{Wang2020}%
  \BibitemOpen
  \bibfield  {author} {\bibinfo {author} {\bibfnamefont {R.}~\bibnamefont
  {Wang}}, \bibinfo {author} {\bibfnamefont {Y.}~\bibnamefont {Su}}, \bibinfo
  {author} {\bibfnamefont {G.}~\bibnamefont {Yang}}, \bibinfo {author}
  {\bibfnamefont {J.}~\bibnamefont {Zhang}}, \ and\ \bibinfo {author}
  {\bibfnamefont {S.}~\bibnamefont {Zhang}},\ }\href {\doibase
  10.1021/acs.chemmater.9b04645} {\bibfield  {journal} {\bibinfo  {journal}
  {Chemistry of Materials}\ }\textbf {\bibinfo {volume} {32}},\ \bibinfo
  {pages} {1545} (\bibinfo {year} {2020})}\BibitemShut {NoStop}%
\bibitem [{\citenamefont {Pizzochero}(2020)}]{Pizzochero2020b}%
  \BibitemOpen
  \bibfield  {author} {\bibinfo {author} {\bibfnamefont {M.}~\bibnamefont
  {Pizzochero}},\ }\href {\doibase 10.1088/1361-6463/ab7ca3} {\bibfield
  {journal} {\bibinfo  {journal} {Journal of Physics D: Applied Physics}\
  }\textbf {\bibinfo {volume} {53}},\ \bibinfo {pages} {244003} (\bibinfo
  {year} {2020})}\BibitemShut {NoStop}%
\bibitem [{\citenamefont {Singamaneni}\ \emph {et~al.}(2020)\citenamefont
  {Singamaneni}, \citenamefont {Martinez}, \citenamefont {Niklas},
  \citenamefont {Poluektov}, \citenamefont {Yadav}, \citenamefont {Pizzochero},
  \citenamefont {Yazyev},\ and\ \citenamefont {McGuire}}]{Singamaneni2020}%
  \BibitemOpen
  \bibfield  {author} {\bibinfo {author} {\bibfnamefont {S.~R.}\ \bibnamefont
  {Singamaneni}}, \bibinfo {author} {\bibfnamefont {L.~M.}\ \bibnamefont
  {Martinez}}, \bibinfo {author} {\bibfnamefont {J.}~\bibnamefont {Niklas}},
  \bibinfo {author} {\bibfnamefont {O.~G.}\ \bibnamefont {Poluektov}}, \bibinfo
  {author} {\bibfnamefont {R.}~\bibnamefont {Yadav}}, \bibinfo {author}
  {\bibfnamefont {M.}~\bibnamefont {Pizzochero}}, \bibinfo {author}
  {\bibfnamefont {O.~V.}\ \bibnamefont {Yazyev}}, \ and\ \bibinfo {author}
  {\bibfnamefont {M.~A.}\ \bibnamefont {McGuire}},\ }\href {\doibase
  10.1063/5.0010888} {\bibfield  {journal} {\bibinfo  {journal} {Applied
  Physics Letters}\ }\textbf {\bibinfo {volume} {117}},\ \bibinfo {pages}
  {082406} (\bibinfo {year} {2020})}\BibitemShut {NoStop}%
\bibitem [{\citenamefont {Jang}\ \emph {et~al.}(2019)\citenamefont {Jang},
  \citenamefont {Jeong}, \citenamefont {Yoon}, \citenamefont {Ryee},\ and\
  \citenamefont {Han}}]{PhysRevMaterials.3.031001}%
  \BibitemOpen
  \bibfield  {author} {\bibinfo {author} {\bibfnamefont {S.~W.}\ \bibnamefont
  {Jang}}, \bibinfo {author} {\bibfnamefont {M.~Y.}\ \bibnamefont {Jeong}},
  \bibinfo {author} {\bibfnamefont {H.}~\bibnamefont {Yoon}}, \bibinfo {author}
  {\bibfnamefont {S.}~\bibnamefont {Ryee}}, \ and\ \bibinfo {author}
  {\bibfnamefont {M.~J.}\ \bibnamefont {Han}},\ }\href {\doibase
  10.1103/PhysRevMaterials.3.031001} {\bibfield  {journal} {\bibinfo  {journal}
  {Phys. Rev. Materials}\ }\textbf {\bibinfo {volume} {3}},\ \bibinfo {pages}
  {031001} (\bibinfo {year} {2019})}\BibitemShut {NoStop}%
\bibitem [{\citenamefont {Besbes}\ \emph {et~al.}(2019)\citenamefont {Besbes},
  \citenamefont {Nikolaev}, \citenamefont {Meskini},\ and\ \citenamefont
  {Solovyev}}]{PhysRevB.99.104432}%
  \BibitemOpen
  \bibfield  {author} {\bibinfo {author} {\bibfnamefont {O.}~\bibnamefont
  {Besbes}}, \bibinfo {author} {\bibfnamefont {S.}~\bibnamefont {Nikolaev}},
  \bibinfo {author} {\bibfnamefont {N.}~\bibnamefont {Meskini}}, \ and\
  \bibinfo {author} {\bibfnamefont {I.}~\bibnamefont {Solovyev}},\ }\href
  {\doibase 10.1103/PhysRevB.99.104432} {\bibfield  {journal} {\bibinfo
  {journal} {Phys. Rev. B}\ }\textbf {\bibinfo {volume} {99}},\ \bibinfo
  {pages} {104432} (\bibinfo {year} {2019})}\BibitemShut {NoStop}%
\bibitem [{\citenamefont {Soriano}\ \emph {et~al.}(2021)\citenamefont
  {Soriano}, \citenamefont {Rudenko}, \citenamefont {Katsnelson},\ and\
  \citenamefont {R\"osner}}]{soriano_environmental_2021}%
  \BibitemOpen
  \bibfield  {author} {\bibinfo {author} {\bibfnamefont {D.}~\bibnamefont
  {Soriano}}, \bibinfo {author} {\bibfnamefont {A.~N.}\ \bibnamefont
  {Rudenko}}, \bibinfo {author} {\bibfnamefont {M.~I.}\ \bibnamefont
  {Katsnelson}}, \ and\ \bibinfo {author} {\bibfnamefont {M.}~\bibnamefont
  {R\"osner}},\ }\href {\doibase 10.1038/s41524-021-00631-4} {\bibfield
  {journal} {\bibinfo  {journal} {npj Computational Materials}\ }\textbf
  {\bibinfo {volume} {7}},\ \bibinfo {pages} {162} (\bibinfo {year}
  {2021})}\BibitemShut {NoStop}%
\bibitem [{\citenamefont {Wang}\ and\ \citenamefont {Sanyal}(2021)}]{Wang21}%
  \BibitemOpen
  \bibfield  {author} {\bibinfo {author} {\bibfnamefont {D.}~\bibnamefont
  {Wang}}\ and\ \bibinfo {author} {\bibfnamefont {B.}~\bibnamefont {Sanyal}},\
  }\href {\doibase 10.1021/acs.jpcc.1c04311} {\bibfield  {journal} {\bibinfo
  {journal} {The Journal of Physical Chemistry C}\ }\textbf {\bibinfo {volume}
  {125}},\ \bibinfo {pages} {18467} (\bibinfo {year} {2021})}\BibitemShut
  {NoStop}%
\bibitem [{\citenamefont {McGuire}\ \emph {et~al.}(2017)\citenamefont
  {McGuire}, \citenamefont {Clark}, \citenamefont {KC}, \citenamefont {Chance},
  \citenamefont {Jellison}, \citenamefont {Cooper}, \citenamefont {Xu},\ and\
  \citenamefont {Sales}}]{McGuire2017}%
  \BibitemOpen
  \bibfield  {author} {\bibinfo {author} {\bibfnamefont {M.~A.}\ \bibnamefont
  {McGuire}}, \bibinfo {author} {\bibfnamefont {G.}~\bibnamefont {Clark}},
  \bibinfo {author} {\bibfnamefont {S.}~\bibnamefont {KC}}, \bibinfo {author}
  {\bibfnamefont {W.~M.}\ \bibnamefont {Chance}}, \bibinfo {author}
  {\bibfnamefont {G.~E.}\ \bibnamefont {Jellison}}, \bibinfo {author}
  {\bibfnamefont {V.~R.}\ \bibnamefont {Cooper}}, \bibinfo {author}
  {\bibfnamefont {X.}~\bibnamefont {Xu}}, \ and\ \bibinfo {author}
  {\bibfnamefont {B.~C.}\ \bibnamefont {Sales}},\ }\href {\doibase
  10.1103/PhysRevMaterials.1.014001} {\bibfield  {journal} {\bibinfo  {journal}
  {Physical Review Materials}\ }\textbf {\bibinfo {volume} {1}},\ \bibinfo
  {pages} {014001} (\bibinfo {year} {2017})}\BibitemShut {NoStop}%
\bibitem [{\citenamefont {Irkhin}\ \emph {et~al.}(1999)\citenamefont {Irkhin},
  \citenamefont {Katanin},\ and\ \citenamefont
  {Katsnelson}}]{PhysRevB.60.1082}%
  \BibitemOpen
  \bibfield  {author} {\bibinfo {author} {\bibfnamefont {V.~Y.}\ \bibnamefont
  {Irkhin}}, \bibinfo {author} {\bibfnamefont {A.~A.}\ \bibnamefont {Katanin}},
  \ and\ \bibinfo {author} {\bibfnamefont {M.~I.}\ \bibnamefont {Katsnelson}},\
  }\href {\doibase 10.1103/PhysRevB.60.1082} {\bibfield  {journal} {\bibinfo
  {journal} {Phys. Rev. B}\ }\textbf {\bibinfo {volume} {60}},\ \bibinfo
  {pages} {1082} (\bibinfo {year} {1999})}\BibitemShut {NoStop}%
\bibitem [{\citenamefont {Chen}\ \emph {et~al.}(2018)\citenamefont {Chen},
  \citenamefont {Chung}, \citenamefont {Gao}, \citenamefont {Chen},
  \citenamefont {Stone}, \citenamefont {Kolesnikov}, \citenamefont {Huang},\
  and\ \citenamefont {Dai}}]{PhysRevX.8.041028}%
  \BibitemOpen
  \bibfield  {author} {\bibinfo {author} {\bibfnamefont {L.}~\bibnamefont
  {Chen}}, \bibinfo {author} {\bibfnamefont {J.-H.}\ \bibnamefont {Chung}},
  \bibinfo {author} {\bibfnamefont {B.}~\bibnamefont {Gao}}, \bibinfo {author}
  {\bibfnamefont {T.}~\bibnamefont {Chen}}, \bibinfo {author} {\bibfnamefont
  {M.~B.}\ \bibnamefont {Stone}}, \bibinfo {author} {\bibfnamefont {A.~I.}\
  \bibnamefont {Kolesnikov}}, \bibinfo {author} {\bibfnamefont
  {Q.}~\bibnamefont {Huang}}, \ and\ \bibinfo {author} {\bibfnamefont
  {P.}~\bibnamefont {Dai}},\ }\href {\doibase 10.1103/PhysRevX.8.041028}
  {\bibfield  {journal} {\bibinfo  {journal} {Phys. Rev. X}\ }\textbf {\bibinfo
  {volume} {8}},\ \bibinfo {pages} {041028} (\bibinfo {year}
  {2018})}\BibitemShut {NoStop}%
\bibitem [{\citenamefont {Lee}\ \emph {et~al.}(2020{\natexlab{a}})\citenamefont
  {Lee}, \citenamefont {Utermohlen}, \citenamefont {Weber}, \citenamefont
  {Hwang}, \citenamefont {Zhang}, \citenamefont {van Tol}, \citenamefont
  {Goldberger}, \citenamefont {Trivedi},\ and\ \citenamefont
  {Hammel}}]{PhysRevLett.124.017201}%
  \BibitemOpen
  \bibfield  {author} {\bibinfo {author} {\bibfnamefont {I.}~\bibnamefont
  {Lee}}, \bibinfo {author} {\bibfnamefont {F.~G.}\ \bibnamefont {Utermohlen}},
  \bibinfo {author} {\bibfnamefont {D.}~\bibnamefont {Weber}}, \bibinfo
  {author} {\bibfnamefont {K.}~\bibnamefont {Hwang}}, \bibinfo {author}
  {\bibfnamefont {C.}~\bibnamefont {Zhang}}, \bibinfo {author} {\bibfnamefont
  {J.}~\bibnamefont {van Tol}}, \bibinfo {author} {\bibfnamefont {J.~E.}\
  \bibnamefont {Goldberger}}, \bibinfo {author} {\bibfnamefont
  {N.}~\bibnamefont {Trivedi}}, \ and\ \bibinfo {author} {\bibfnamefont
  {P.~C.}\ \bibnamefont {Hammel}},\ }\href {\doibase
  10.1103/PhysRevLett.124.017201} {\bibfield  {journal} {\bibinfo  {journal}
  {Phys. Rev. Lett.}\ }\textbf {\bibinfo {volume} {124}},\ \bibinfo {pages}
  {017201} (\bibinfo {year} {2020}{\natexlab{a}})}\BibitemShut {NoStop}%
\bibitem [{\citenamefont {Kvashnin}\ \emph {et~al.}(2020)\citenamefont
  {Kvashnin}, \citenamefont {Bergman}, \citenamefont {Lichtenstein},\ and\
  \citenamefont {Katsnelson}}]{PhysRevB.102.115162}%
  \BibitemOpen
  \bibfield  {author} {\bibinfo {author} {\bibfnamefont {Y.~O.}\ \bibnamefont
  {Kvashnin}}, \bibinfo {author} {\bibfnamefont {A.}~\bibnamefont {Bergman}},
  \bibinfo {author} {\bibfnamefont {A.~I.}\ \bibnamefont {Lichtenstein}}, \
  and\ \bibinfo {author} {\bibfnamefont {M.~I.}\ \bibnamefont {Katsnelson}},\
  }\href {\doibase 10.1103/PhysRevB.102.115162} {\bibfield  {journal} {\bibinfo
   {journal} {Phys. Rev. B}\ }\textbf {\bibinfo {volume} {102}},\ \bibinfo
  {pages} {115162} (\bibinfo {year} {2020})}\BibitemShut {NoStop}%
\bibitem [{\citenamefont {Ke}\ and\ \citenamefont
  {Katsnelson}(2021)}]{ke_electron_2021}%
  \BibitemOpen
  \bibfield  {author} {\bibinfo {author} {\bibfnamefont {L.}~\bibnamefont
  {Ke}}\ and\ \bibinfo {author} {\bibfnamefont {M.~I.}\ \bibnamefont
  {Katsnelson}},\ }\href {\doibase 10.1038/s41524-020-00469-2} {\bibfield
  {journal} {\bibinfo  {journal} {npj Computational Materials}\ }\textbf
  {\bibinfo {volume} {7}},\ \bibinfo {pages} {4} (\bibinfo {year}
  {2021})}\BibitemShut {NoStop}%
\bibitem [{\citenamefont {Kartsev}\ \emph {et~al.}(2020)\citenamefont
  {Kartsev}, \citenamefont {Augustin}, \citenamefont {Evans}, \citenamefont
  {Novoselov},\ and\ \citenamefont {Santos}}]{Kartsev2020}%
  \BibitemOpen
  \bibfield  {author} {\bibinfo {author} {\bibfnamefont {A.}~\bibnamefont
  {Kartsev}}, \bibinfo {author} {\bibfnamefont {M.}~\bibnamefont {Augustin}},
  \bibinfo {author} {\bibfnamefont {R.~F.~L.}\ \bibnamefont {Evans}}, \bibinfo
  {author} {\bibfnamefont {K.~S.}\ \bibnamefont {Novoselov}}, \ and\ \bibinfo
  {author} {\bibfnamefont {E.~J.~G.}\ \bibnamefont {Santos}},\ }\href {\doibase
  10.1038/s41524-020-00416-1} {\bibfield  {journal} {\bibinfo  {journal} {npj
  Computational Materials}\ }\textbf {\bibinfo {volume} {6}},\ \bibinfo {pages}
  {150} (\bibinfo {year} {2020})}\BibitemShut {NoStop}%
\bibitem [{\citenamefont {Wahab}\ \emph {et~al.}(2021)\citenamefont {Wahab},
  \citenamefont {Augustin}, \citenamefont {Valero}, \citenamefont {Kuang},
  \citenamefont {Jenkins}, \citenamefont {Coronado}, \citenamefont
  {Grigorieva}, \citenamefont {Vera-Marun}, \citenamefont {Navarro-Moratalla},
  \citenamefont {Evans}, \citenamefont {Novoselov},\ and\ \citenamefont
  {Santos}}]{Wahab2021}%
  \BibitemOpen
  \bibfield  {author} {\bibinfo {author} {\bibfnamefont {D.~A.}\ \bibnamefont
  {Wahab}}, \bibinfo {author} {\bibfnamefont {M.}~\bibnamefont {Augustin}},
  \bibinfo {author} {\bibfnamefont {S.~M.}\ \bibnamefont {Valero}}, \bibinfo
  {author} {\bibfnamefont {W.}~\bibnamefont {Kuang}}, \bibinfo {author}
  {\bibfnamefont {S.}~\bibnamefont {Jenkins}}, \bibinfo {author} {\bibfnamefont
  {E.}~\bibnamefont {Coronado}}, \bibinfo {author} {\bibfnamefont {I.~V.}\
  \bibnamefont {Grigorieva}}, \bibinfo {author} {\bibfnamefont {I.~J.}\
  \bibnamefont {Vera-Marun}}, \bibinfo {author} {\bibfnamefont
  {E.}~\bibnamefont {Navarro-Moratalla}}, \bibinfo {author} {\bibfnamefont
  {R.~F.~L.}\ \bibnamefont {Evans}}, \bibinfo {author} {\bibfnamefont {K.~S.}\
  \bibnamefont {Novoselov}}, \ and\ \bibinfo {author} {\bibfnamefont
  {E.~J.~G.}\ \bibnamefont {Santos}},\ }\href {\doibase
  https://doi.org/10.1002/adma.202004138} {\bibfield  {journal} {\bibinfo
  {journal} {Advanced Materials}\ }\textbf {\bibinfo {volume} {33}},\ \bibinfo
  {pages} {2004138} (\bibinfo {year} {2021})}\BibitemShut {NoStop}%
\bibitem [{\citenamefont {Shao}\ \emph {et~al.}(2021)\citenamefont {Shao},
  \citenamefont {Karki}, \citenamefont {Huang}, \citenamefont {Feng},
  \citenamefont {Sumanasekera}, \citenamefont {Guo}, \citenamefont {Chuang},\
  and\ \citenamefont {Freelon}}]{Shao2021}%
  \BibitemOpen
  \bibfield  {author} {\bibinfo {author} {\bibfnamefont {Y.~C.}\ \bibnamefont
  {Shao}}, \bibinfo {author} {\bibfnamefont {B.}~\bibnamefont {Karki}},
  \bibinfo {author} {\bibfnamefont {W.}~\bibnamefont {Huang}}, \bibinfo
  {author} {\bibfnamefont {X.}~\bibnamefont {Feng}}, \bibinfo {author}
  {\bibfnamefont {G.}~\bibnamefont {Sumanasekera}}, \bibinfo {author}
  {\bibfnamefont {J.-H.}\ \bibnamefont {Guo}}, \bibinfo {author} {\bibfnamefont
  {Y.-D.}\ \bibnamefont {Chuang}}, \ and\ \bibinfo {author} {\bibfnamefont
  {B.}~\bibnamefont {Freelon}},\ }\href {\doibase 10.1021/acs.jpclett.0c03476}
  {\bibfield  {journal} {\bibinfo  {journal} {The Journal of Physical Chemistry
  Letters}\ }\textbf {\bibinfo {volume} {12}},\ \bibinfo {pages} {724}
  (\bibinfo {year} {2021})}\BibitemShut {NoStop}%
\bibitem [{\citenamefont {Lee}\ \emph {et~al.}(2020{\natexlab{b}})\citenamefont
  {Lee}, \citenamefont {Utermohlen}, \citenamefont {Weber}, \citenamefont
  {Hwang}, \citenamefont {Zhang}, \citenamefont {van Tol}, \citenamefont
  {Goldberger}, \citenamefont {Trivedi},\ and\ \citenamefont
  {Hammel}}]{Lee2020}%
  \BibitemOpen
  \bibfield  {author} {\bibinfo {author} {\bibfnamefont {I.}~\bibnamefont
  {Lee}}, \bibinfo {author} {\bibfnamefont {F.~G.}\ \bibnamefont {Utermohlen}},
  \bibinfo {author} {\bibfnamefont {D.}~\bibnamefont {Weber}}, \bibinfo
  {author} {\bibfnamefont {K.}~\bibnamefont {Hwang}}, \bibinfo {author}
  {\bibfnamefont {C.}~\bibnamefont {Zhang}}, \bibinfo {author} {\bibfnamefont
  {J.}~\bibnamefont {van Tol}}, \bibinfo {author} {\bibfnamefont {J.~E.}\
  \bibnamefont {Goldberger}}, \bibinfo {author} {\bibfnamefont
  {N.}~\bibnamefont {Trivedi}}, \ and\ \bibinfo {author} {\bibfnamefont
  {P.~C.}\ \bibnamefont {Hammel}},\ }\href {\doibase
  10.1103/PhysRevLett.124.017201} {\bibfield  {journal} {\bibinfo  {journal}
  {Physical Review Letters}\ }\textbf {\bibinfo {volume} {124}},\ \bibinfo
  {pages} {017201} (\bibinfo {year} {2020}{\natexlab{b}})}\BibitemShut
  {NoStop}%
\bibitem [{\citenamefont {Cenker}\ \emph {et~al.}(2021)\citenamefont {Cenker},
  \citenamefont {Huang}, \citenamefont {Suri}, \citenamefont {Thijssen},
  \citenamefont {Miller}, \citenamefont {Song}, \citenamefont {Taniguchi},
  \citenamefont {Watanabe}, \citenamefont {McGuire}, \citenamefont {Xiao},\
  and\ \citenamefont {Xu}}]{Cenker2021}%
  \BibitemOpen
  \bibfield  {author} {\bibinfo {author} {\bibfnamefont {J.}~\bibnamefont
  {Cenker}}, \bibinfo {author} {\bibfnamefont {B.}~\bibnamefont {Huang}},
  \bibinfo {author} {\bibfnamefont {N.}~\bibnamefont {Suri}}, \bibinfo {author}
  {\bibfnamefont {P.}~\bibnamefont {Thijssen}}, \bibinfo {author}
  {\bibfnamefont {A.}~\bibnamefont {Miller}}, \bibinfo {author} {\bibfnamefont
  {T.}~\bibnamefont {Song}}, \bibinfo {author} {\bibfnamefont {T.}~\bibnamefont
  {Taniguchi}}, \bibinfo {author} {\bibfnamefont {K.}~\bibnamefont {Watanabe}},
  \bibinfo {author} {\bibfnamefont {M.~A.}\ \bibnamefont {McGuire}}, \bibinfo
  {author} {\bibfnamefont {D.}~\bibnamefont {Xiao}}, \ and\ \bibinfo {author}
  {\bibfnamefont {X.}~\bibnamefont {Xu}},\ }\href {\doibase
  10.1038/s41567-020-0999-1} {\bibfield  {journal} {\bibinfo  {journal} {Nature
  Physics}\ }\textbf {\bibinfo {volume} {17}},\ \bibinfo {pages} {20} (\bibinfo
  {year} {2021})}\BibitemShut {NoStop}%
\bibitem [{\citenamefont {Pizzochero}\ \emph {et~al.}(2020)\citenamefont
  {Pizzochero}, \citenamefont {Yadav},\ and\ \citenamefont
  {Yazyev}}]{Pizzochero2020c}%
  \BibitemOpen
  \bibfield  {author} {\bibinfo {author} {\bibfnamefont {M.}~\bibnamefont
  {Pizzochero}}, \bibinfo {author} {\bibfnamefont {R.}~\bibnamefont {Yadav}}, \
  and\ \bibinfo {author} {\bibfnamefont {O.~V.}\ \bibnamefont {Yazyev}},\
  }\href {\doibase 10.1088/2053-1583/ab7cab} {\bibfield  {journal} {\bibinfo
  {journal} {2D Materials}\ }\textbf {\bibinfo {volume} {7}},\ \bibinfo {pages}
  {035005} (\bibinfo {year} {2020})}\BibitemShut {NoStop}%
\bibitem [{\citenamefont {Menichetti}\ \emph {et~al.}(2019)\citenamefont
  {Menichetti}, \citenamefont {Calandra},\ and\ \citenamefont
  {Polini}}]{Menichetti2019}%
  \BibitemOpen
  \bibfield  {author} {\bibinfo {author} {\bibfnamefont {G.}~\bibnamefont
  {Menichetti}}, \bibinfo {author} {\bibfnamefont {M.}~\bibnamefont
  {Calandra}}, \ and\ \bibinfo {author} {\bibfnamefont {M.}~\bibnamefont
  {Polini}},\ }\href {\doibase 10.1088/2053-1583/ab2f06} {\bibfield  {journal}
  {\bibinfo  {journal} {2D Materials}\ }\textbf {\bibinfo {volume} {6}},\
  \bibinfo {pages} {045042} (\bibinfo {year} {2019})}\BibitemShut {NoStop}%
\bibitem [{\citenamefont {Klinkova}\ and\ \citenamefont
  {Bochkareva}(1980)}]{Klinkova1980}%
  \BibitemOpen
  \bibfield  {author} {\bibinfo {author} {\bibfnamefont {L.}~\bibnamefont
  {Klinkova}}\ and\ \bibinfo {author} {\bibfnamefont {V.}~\bibnamefont
  {Bochkareva}},\ }\href@noop {} {\bibfield  {journal} {\bibinfo  {journal}
  {Inorganic Materials}\ }\textbf {\bibinfo {volume} {16}},\ \bibinfo {pages}
  {1207} (\bibinfo {year} {1980})}\BibitemShut {NoStop}%
\bibitem [{\citenamefont {McGuire}\ \emph {et~al.}(2015)\citenamefont
  {McGuire}, \citenamefont {Dixit}, \citenamefont {Cooper},\ and\ \citenamefont
  {Sales}}]{McGuire2015}%
  \BibitemOpen
  \bibfield  {author} {\bibinfo {author} {\bibfnamefont {M.~A.}\ \bibnamefont
  {McGuire}}, \bibinfo {author} {\bibfnamefont {H.}~\bibnamefont {Dixit}},
  \bibinfo {author} {\bibfnamefont {V.~R.}\ \bibnamefont {Cooper}}, \ and\
  \bibinfo {author} {\bibfnamefont {B.~C.}\ \bibnamefont {Sales}},\ }\href
  {\doibase 10.1021/cm504242t} {\bibfield  {journal} {\bibinfo  {journal}
  {Chemistry of Materials}\ }\textbf {\bibinfo {volume} {27}},\ \bibinfo
  {pages} {612} (\bibinfo {year} {2015})}\BibitemShut {NoStop}%
\bibitem [{\citenamefont {Klintenberg}\ \emph {et~al.}(2000)\citenamefont
  {Klintenberg}, \citenamefont {Derenzo},\ and\ \citenamefont
  {Weber}}]{Klintenberg2000}%
  \BibitemOpen
  \bibfield  {author} {\bibinfo {author} {\bibfnamefont {M.}~\bibnamefont
  {Klintenberg}}, \bibinfo {author} {\bibfnamefont {S.}~\bibnamefont
  {Derenzo}}, \ and\ \bibinfo {author} {\bibfnamefont {M.}~\bibnamefont
  {Weber}},\ }\href {\doibase https://doi.org/10.1016/S0010-4655(00)00071-0}
  {\bibfield  {journal} {\bibinfo  {journal} {Computer Physics Communications}\
  }\textbf {\bibinfo {volume} {131}},\ \bibinfo {pages} {120} (\bibinfo {year}
  {2000})}\BibitemShut {NoStop}%
\bibitem [{\citenamefont {Helgaker}\ \emph {et~al.}(2000)\citenamefont
  {Helgaker}, \citenamefont {J{\o}rgensen},\ and\ \citenamefont
  {Olsen}}]{Helgaker2000}%
  \BibitemOpen
  \bibfield  {author} {\bibinfo {author} {\bibfnamefont {T.}~\bibnamefont
  {Helgaker}}, \bibinfo {author} {\bibfnamefont {P.}~\bibnamefont
  {J{\o}rgensen}}, \ and\ \bibinfo {author} {\bibfnamefont {J.}~\bibnamefont
  {Olsen}},\ }\href@noop {} {\emph {\bibinfo {title} {Molecular
  Electronic-Structure Theory}}}\ (\bibinfo  {publisher} {Wiley, Chichester},\
  \bibinfo {year} {2000})\BibitemShut {NoStop}%
\bibitem [{\citenamefont {Moreira}\ and\ \citenamefont
  {Illas}(2006)}]{Moreira2006}%
  \BibitemOpen
  \bibfield  {author} {\bibinfo {author} {\bibfnamefont {I.~d. P.~R.}\
  \bibnamefont {Moreira}}\ and\ \bibinfo {author} {\bibfnamefont
  {F.}~\bibnamefont {Illas}},\ }\href {\doibase 10.1039/B515732C} {\bibfield
  {journal} {\bibinfo  {journal} {Physical Chemistry Chemical Physics}\
  }\textbf {\bibinfo {volume} {8}},\ \bibinfo {pages} {1645} (\bibinfo {year}
  {2006})}\BibitemShut {NoStop}%
\bibitem [{\citenamefont {Malrieu}\ \emph {et~al.}(2014)\citenamefont
  {Malrieu}, \citenamefont {Caballol}, \citenamefont {Calzado}, \citenamefont
  {de~Graaf},\ and\ \citenamefont {Guih\'ery}}]{Malrieu2014}%
  \BibitemOpen
  \bibfield  {author} {\bibinfo {author} {\bibfnamefont {J.~P.}\ \bibnamefont
  {Malrieu}}, \bibinfo {author} {\bibfnamefont {R.}~\bibnamefont {Caballol}},
  \bibinfo {author} {\bibfnamefont {C.~J.}\ \bibnamefont {Calzado}}, \bibinfo
  {author} {\bibfnamefont {C.}~\bibnamefont {de~Graaf}}, \ and\ \bibinfo
  {author} {\bibfnamefont {N.}~\bibnamefont {Guih\'ery}},\ }\href {\doibase
  10.1021/cr300500z} {\bibfield  {journal} {\bibinfo  {journal} {Chemical
  Reviews}\ }\textbf {\bibinfo {volume} {114}},\ \bibinfo {pages} {429}
  (\bibinfo {year} {2014})}\BibitemShut {NoStop}%
\bibitem [{\citenamefont {Bogdanov}\ \emph {et~al.}(2011)\citenamefont
  {Bogdanov}, \citenamefont {van~den Brink},\ and\ \citenamefont
  {Hozoi}}]{Bogdanov2011}%
  \BibitemOpen
  \bibfield  {author} {\bibinfo {author} {\bibfnamefont {N.~A.}\ \bibnamefont
  {Bogdanov}}, \bibinfo {author} {\bibfnamefont {J.}~\bibnamefont {van~den
  Brink}}, \ and\ \bibinfo {author} {\bibfnamefont {L.}~\bibnamefont {Hozoi}},\
  }\href {\doibase 10.1103/PhysRevB.84.235146} {\bibfield  {journal} {\bibinfo
  {journal} {Physical Review B}\ }\textbf {\bibinfo {volume} {84}},\ \bibinfo
  {pages} {235146} (\bibinfo {year} {2011})}\BibitemShut {NoStop}%
\bibitem [{\citenamefont {Katukuri}\ \emph {et~al.}(2012)\citenamefont
  {Katukuri}, \citenamefont {Stoll}, \citenamefont {van~den Brink},\ and\
  \citenamefont {Hozoi}}]{Katukuri2012}%
  \BibitemOpen
  \bibfield  {author} {\bibinfo {author} {\bibfnamefont {V.~M.}\ \bibnamefont
  {Katukuri}}, \bibinfo {author} {\bibfnamefont {H.}~\bibnamefont {Stoll}},
  \bibinfo {author} {\bibfnamefont {J.}~\bibnamefont {van~den Brink}}, \ and\
  \bibinfo {author} {\bibfnamefont {L.}~\bibnamefont {Hozoi}},\ }\href
  {\doibase 10.1103/PhysRevB.85.220402} {\bibfield  {journal} {\bibinfo
  {journal} {Physical Review B}\ }\textbf {\bibinfo {volume} {85}},\ \bibinfo
  {pages} {220402} (\bibinfo {year} {2012})}\BibitemShut {NoStop}%
\bibitem [{\citenamefont {Moreira}\ \emph {et~al.}(1999)\citenamefont
  {Moreira}, \citenamefont {Illas}, \citenamefont {Calzado}, \citenamefont
  {Sanz}, \citenamefont {Malrieu}, \citenamefont {Amor},\ and\ \citenamefont
  {Maynau}}]{Moreira1999}%
  \BibitemOpen
  \bibfield  {author} {\bibinfo {author} {\bibfnamefont {I.~d. P.~R.}\
  \bibnamefont {Moreira}}, \bibinfo {author} {\bibfnamefont {F.}~\bibnamefont
  {Illas}}, \bibinfo {author} {\bibfnamefont {C.~J.}\ \bibnamefont {Calzado}},
  \bibinfo {author} {\bibfnamefont {J.~F.}\ \bibnamefont {Sanz}}, \bibinfo
  {author} {\bibfnamefont {J.-P.}\ \bibnamefont {Malrieu}}, \bibinfo {author}
  {\bibfnamefont {N.~B.}\ \bibnamefont {Amor}}, \ and\ \bibinfo {author}
  {\bibfnamefont {D.}~\bibnamefont {Maynau}},\ }\href {\doibase
  10.1103/PhysRevB.59.R6593} {\bibfield  {journal} {\bibinfo  {journal}
  {Physical Review B}\ }\textbf {\bibinfo {volume} {59}},\ \bibinfo {pages}
  {R6593} (\bibinfo {year} {1999})}\BibitemShut {NoStop}%
\bibitem [{\citenamefont {Bogdanov}\ \emph {et~al.}(2013)\citenamefont
  {Bogdanov}, \citenamefont {Maurice}, \citenamefont {Rousochatzakis},
  \citenamefont {van~den Brink},\ and\ \citenamefont {Hozoi}}]{Bogdanov2013}%
  \BibitemOpen
  \bibfield  {author} {\bibinfo {author} {\bibfnamefont {N.~A.}\ \bibnamefont
  {Bogdanov}}, \bibinfo {author} {\bibfnamefont {R.}~\bibnamefont {Maurice}},
  \bibinfo {author} {\bibfnamefont {I.}~\bibnamefont {Rousochatzakis}},
  \bibinfo {author} {\bibfnamefont {J.}~\bibnamefont {van~den Brink}}, \ and\
  \bibinfo {author} {\bibfnamefont {L.}~\bibnamefont {Hozoi}},\ }\href
  {\doibase 10.1103/PhysRevLett.110.127206} {\bibfield  {journal} {\bibinfo
  {journal} {Physical Review Letters}\ }\textbf {\bibinfo {volume} {110}},\
  \bibinfo {pages} {127206} (\bibinfo {year} {2013})}\BibitemShut {NoStop}%
\bibitem [{\citenamefont {Bogdanov}\ \emph {et~al.}(2017)\citenamefont
  {Bogdanov}, \citenamefont {Bisogni}, \citenamefont {Kraus}, \citenamefont
  {Monney}, \citenamefont {Zhou}, \citenamefont {Schmitt}, \citenamefont
  {Geck}, \citenamefont {Mitrushchenkov}, \citenamefont {Stoll}, \citenamefont
  {van~den Brink},\ and\ \citenamefont {Hozoi}}]{Nikolary_d9_rixs}%
  \BibitemOpen
  \bibfield  {author} {\bibinfo {author} {\bibfnamefont {N.~A.}\ \bibnamefont
  {Bogdanov}}, \bibinfo {author} {\bibfnamefont {V.}~\bibnamefont {Bisogni}},
  \bibinfo {author} {\bibfnamefont {R.}~\bibnamefont {Kraus}}, \bibinfo
  {author} {\bibfnamefont {C.}~\bibnamefont {Monney}}, \bibinfo {author}
  {\bibfnamefont {K.}~\bibnamefont {Zhou}}, \bibinfo {author} {\bibfnamefont
  {T.}~\bibnamefont {Schmitt}}, \bibinfo {author} {\bibfnamefont
  {J.}~\bibnamefont {Geck}}, \bibinfo {author} {\bibfnamefont {A.~O.}\
  \bibnamefont {Mitrushchenkov}}, \bibinfo {author} {\bibfnamefont
  {H.}~\bibnamefont {Stoll}}, \bibinfo {author} {\bibfnamefont
  {J.}~\bibnamefont {van~den Brink}}, \ and\ \bibinfo {author} {\bibfnamefont
  {L.}~\bibnamefont {Hozoi}},\ }\href@noop {} {\bibfield  {journal} {\bibinfo
  {journal} {Journal of Physics: Condensed Matter}\ }\textbf {\bibinfo {volume}
  {29}},\ \bibinfo {pages} {035502} (\bibinfo {year} {2017})}\BibitemShut
  {NoStop}%
\bibitem [{\citenamefont {Li}\ \emph {et~al.}(2021)\citenamefont {Li},
  \citenamefont {Xu}, \citenamefont {Garcia-Fernandez}, \citenamefont {Nag},
  \citenamefont {Robarts}, \citenamefont {Walters}, \citenamefont {Liu},
  \citenamefont {Zhou}, \citenamefont {Wohlfeld}, \citenamefont {van~den
  Brink}, \citenamefont {Ding},\ and\ \citenamefont {Zhou}}]{Li_2021}%
  \BibitemOpen
  \bibfield  {author} {\bibinfo {author} {\bibfnamefont {J.}~\bibnamefont
  {Li}}, \bibinfo {author} {\bibfnamefont {L.}~\bibnamefont {Xu}}, \bibinfo
  {author} {\bibfnamefont {M.}~\bibnamefont {Garcia-Fernandez}}, \bibinfo
  {author} {\bibfnamefont {A.}~\bibnamefont {Nag}}, \bibinfo {author}
  {\bibfnamefont {H.~C.}\ \bibnamefont {Robarts}}, \bibinfo {author}
  {\bibfnamefont {A.~C.}\ \bibnamefont {Walters}}, \bibinfo {author}
  {\bibfnamefont {X.}~\bibnamefont {Liu}}, \bibinfo {author} {\bibfnamefont
  {J.}~\bibnamefont {Zhou}}, \bibinfo {author} {\bibfnamefont {K.}~\bibnamefont
  {Wohlfeld}}, \bibinfo {author} {\bibfnamefont {J.}~\bibnamefont {van~den
  Brink}}, \bibinfo {author} {\bibfnamefont {H.}~\bibnamefont {Ding}}, \ and\
  \bibinfo {author} {\bibfnamefont {K.-J.}\ \bibnamefont {Zhou}},\ }\href
  {\doibase 10.1103/PhysRevLett.126.106401} {\bibfield  {journal} {\bibinfo
  {journal} {Physical Review Letters}\ }\textbf {\bibinfo {volume} {126}},\
  \bibinfo {pages} {106401} (\bibinfo {year} {2021})}\BibitemShut {NoStop}%
\bibitem [{\citenamefont {Fabbris}\ \emph {et~al.}(2017)\citenamefont
  {Fabbris}, \citenamefont {Meyers}, \citenamefont {Xu}, \citenamefont
  {Katukuri}, \citenamefont {Hozoi}, \citenamefont {Liu}, \citenamefont {Chen},
  \citenamefont {Okamoto}, \citenamefont {Schmitt}, \citenamefont {Uldry},
  \citenamefont {Delley}, \citenamefont {Gu}, \citenamefont {Prabhakaran},
  \citenamefont {Boothroyd}, \citenamefont {van~den Brink}, \citenamefont
  {Huang},\ and\ \citenamefont {Dean}}]{Fabbris_2017}%
  \BibitemOpen
  \bibfield  {author} {\bibinfo {author} {\bibfnamefont {G.}~\bibnamefont
  {Fabbris}}, \bibinfo {author} {\bibfnamefont {D.}~\bibnamefont {Meyers}},
  \bibinfo {author} {\bibfnamefont {L.}~\bibnamefont {Xu}}, \bibinfo {author}
  {\bibfnamefont {V.~M.}\ \bibnamefont {Katukuri}}, \bibinfo {author}
  {\bibfnamefont {L.}~\bibnamefont {Hozoi}}, \bibinfo {author} {\bibfnamefont
  {X.}~\bibnamefont {Liu}}, \bibinfo {author} {\bibfnamefont {Z.-Y.}\
  \bibnamefont {Chen}}, \bibinfo {author} {\bibfnamefont {J.}~\bibnamefont
  {Okamoto}}, \bibinfo {author} {\bibfnamefont {T.}~\bibnamefont {Schmitt}},
  \bibinfo {author} {\bibfnamefont {A.}~\bibnamefont {Uldry}}, \bibinfo
  {author} {\bibfnamefont {B.}~\bibnamefont {Delley}}, \bibinfo {author}
  {\bibfnamefont {G.~D.}\ \bibnamefont {Gu}}, \bibinfo {author} {\bibfnamefont
  {D.}~\bibnamefont {Prabhakaran}}, \bibinfo {author} {\bibfnamefont {A.~T.}\
  \bibnamefont {Boothroyd}}, \bibinfo {author} {\bibfnamefont {J.}~\bibnamefont
  {van~den Brink}}, \bibinfo {author} {\bibfnamefont {D.~J.}\ \bibnamefont
  {Huang}}, \ and\ \bibinfo {author} {\bibfnamefont {M.~P.~M.}\ \bibnamefont
  {Dean}},\ }\href {\doibase 10.1103/PhysRevLett.118.156402} {\bibfield
  {journal} {\bibinfo  {journal} {Physical Review Letters}\ }\textbf {\bibinfo
  {volume} {118}},\ \bibinfo {pages} {156402} (\bibinfo {year}
  {2017})}\BibitemShut {NoStop}%
\bibitem [{\citenamefont {van~der Laan}\ and\ \citenamefont
  {Kirkman}(1992)}]{Laan_1992}%
  \BibitemOpen
  \bibfield  {author} {\bibinfo {author} {\bibfnamefont {G.}~\bibnamefont
  {van~der Laan}}\ and\ \bibinfo {author} {\bibfnamefont {I.}~\bibnamefont
  {Kirkman}},\ }\href {\doibase 10.1088/0953-8984/4/16/019} {\bibfield
  {journal} {\bibinfo  {journal} {Journal of Physics: Condensed Matter}\
  }\textbf {\bibinfo {volume} {4}},\ \bibinfo {pages} {4189} (\bibinfo {year}
  {1992})}\BibitemShut {NoStop}%
\bibitem [{\citenamefont {Moriya}(1960)}]{Moriya1960}%
  \BibitemOpen
  \bibfield  {author} {\bibinfo {author} {\bibfnamefont {T.}~\bibnamefont
  {Moriya}},\ }\href {\doibase 10.1103/PhysRev.120.91} {\bibfield  {journal}
  {\bibinfo  {journal} {Physical Review}\ }\textbf {\bibinfo {volume} {120}},\
  \bibinfo {pages} {91} (\bibinfo {year} {1960})}\BibitemShut {NoStop}%
\bibitem [{\citenamefont {Yadav}\ \emph {et~al.}(2018)\citenamefont {Yadav},
  \citenamefont {Ray}, \citenamefont {Eldeeb}, \citenamefont {Nishimoto},
  \citenamefont {Hozoi},\ and\ \citenamefont {van~den Brink}}]{Yadav_2018}%
  \BibitemOpen
  \bibfield  {author} {\bibinfo {author} {\bibfnamefont {R.}~\bibnamefont
  {Yadav}}, \bibinfo {author} {\bibfnamefont {R.}~\bibnamefont {Ray}}, \bibinfo
  {author} {\bibfnamefont {M.~S.}\ \bibnamefont {Eldeeb}}, \bibinfo {author}
  {\bibfnamefont {S.}~\bibnamefont {Nishimoto}}, \bibinfo {author}
  {\bibfnamefont {L.}~\bibnamefont {Hozoi}}, \ and\ \bibinfo {author}
  {\bibfnamefont {J.}~\bibnamefont {van~den Brink}},\ }\href {\doibase
  10.1103/PhysRevLett.121.197203} {\bibfield  {journal} {\bibinfo  {journal}
  {Physical Review Letters}\ }\textbf {\bibinfo {volume} {121}},\ \bibinfo
  {pages} {197203} (\bibinfo {year} {2018})}\BibitemShut {NoStop}%
\bibitem [{\citenamefont {Yadav}\ \emph {et~al.}(2016)\citenamefont {Yadav},
  \citenamefont {Bogdanov}, \citenamefont {Katukuri}, \citenamefont
  {Nishimoto}, \citenamefont {van~den Brink},\ and\ \citenamefont
  {Hozoi}}]{Yadav_2016}%
  \BibitemOpen
  \bibfield  {author} {\bibinfo {author} {\bibfnamefont {R.}~\bibnamefont
  {Yadav}}, \bibinfo {author} {\bibfnamefont {N.~A.}\ \bibnamefont {Bogdanov}},
  \bibinfo {author} {\bibfnamefont {V.~M.}\ \bibnamefont {Katukuri}}, \bibinfo
  {author} {\bibfnamefont {S.}~\bibnamefont {Nishimoto}}, \bibinfo {author}
  {\bibfnamefont {J.}~\bibnamefont {van~den Brink}}, \ and\ \bibinfo {author}
  {\bibfnamefont {L.}~\bibnamefont {Hozoi}},\ }\href {\doibase
  10.1038/srep37925} {\bibfield  {journal} {\bibinfo  {journal} {Scientific
  Reports}\ }\textbf {\bibinfo {volume} {6}},\ \bibinfo {pages} {37925}
  (\bibinfo {year} {2016})}\BibitemShut {NoStop}%
\bibitem [{\citenamefont {Maurice}\ \emph {et~al.}(2009)\citenamefont
  {Maurice}, \citenamefont {Bastardis}, \citenamefont {Graaf}, \citenamefont
  {Suaud}, \citenamefont {Mallah},\ and\ \citenamefont
  {Guih\'ery}}]{Maurice2009}%
  \BibitemOpen
  \bibfield  {author} {\bibinfo {author} {\bibfnamefont {R.}~\bibnamefont
  {Maurice}}, \bibinfo {author} {\bibfnamefont {R.}~\bibnamefont {Bastardis}},
  \bibinfo {author} {\bibfnamefont {C.~d.}\ \bibnamefont {Graaf}}, \bibinfo
  {author} {\bibfnamefont {N.}~\bibnamefont {Suaud}}, \bibinfo {author}
  {\bibfnamefont {T.}~\bibnamefont {Mallah}}, \ and\ \bibinfo {author}
  {\bibfnamefont {N.}~\bibnamefont {Guih\'ery}},\ }\href {\doibase
  10.1021/ct900326e} {\bibfield  {journal} {\bibinfo  {journal} {Journal of
  Chemical Theory and Computation}\ }\textbf {\bibinfo {volume} {5}},\ \bibinfo
  {pages} {2977} (\bibinfo {year} {2009})}\BibitemShut {NoStop}%
\bibitem [{\citenamefont {Lado}\ and\ \citenamefont
  {Fern{\'{a}}ndez-Rossier}(2017)}]{Lado2017}%
  \BibitemOpen
  \bibfield  {author} {\bibinfo {author} {\bibfnamefont {J.~L.}\ \bibnamefont
  {Lado}}\ and\ \bibinfo {author} {\bibfnamefont {J.}~\bibnamefont
  {Fern{\'{a}}ndez-Rossier}},\ }\href {\doibase 10.1088/2053-1583/aa75ed}
  {\bibfield  {journal} {\bibinfo  {journal} {2D Materials}\ }\textbf {\bibinfo
  {volume} {4}},\ \bibinfo {pages} {035002} (\bibinfo {year}
  {2017})}\BibitemShut {NoStop}%
\bibitem [{\citenamefont {Akhiezer}\ \emph {et~al.}(1968)\citenamefont
  {Akhiezer}, \citenamefont {Bar’yakhtar},\ and\ \citenamefont
  {Peletminskii}}]{akhiezer1968spin}%
  \BibitemOpen
  \bibfield  {author} {\bibinfo {author} {\bibfnamefont {A.~I.}\ \bibnamefont
  {Akhiezer}}, \bibinfo {author} {\bibfnamefont {V.~G.}\ \bibnamefont
  {Bar’yakhtar}}, \ and\ \bibinfo {author} {\bibfnamefont {S.~V.}\
  \bibnamefont {Peletminskii}},\ }\href@noop {} {\emph {\bibinfo {title} {{Spin
  Waves}}}}\ (\bibinfo  {publisher} {North-Holland, Amsterdam},\ \bibinfo
  {year} {1968})\BibitemShut {NoStop}%
\bibitem [{\citenamefont {Aharoni}(2000)}]{aharoni2000introduction}%
  \BibitemOpen
  \bibfield  {author} {\bibinfo {author} {\bibfnamefont {A.}~\bibnamefont
  {Aharoni}},\ }\href@noop {} {\emph {\bibinfo {title} {{Introduction to the
  Theory of Ferromagnetism}}}}\ (\bibinfo  {publisher} {Clarendon Press},\
  \bibinfo {year} {2000})\BibitemShut {NoStop}%
\bibitem [{\citenamefont {Kim}\ and\ \citenamefont {Park}(2021)}]{Kim2021}%
  \BibitemOpen
  \bibfield  {author} {\bibinfo {author} {\bibfnamefont {T.~Y.}\ \bibnamefont
  {Kim}}\ and\ \bibinfo {author} {\bibfnamefont {C.-H.}\ \bibnamefont {Park}},\
  }\href {\doibase 10.1021/acs.nanolett.1c03992} {\bibfield  {journal}
  {\bibinfo  {journal} {Nano Lett.}\ }\textbf {\bibinfo {volume} {21}},\
  \bibinfo {pages} {10114} (\bibinfo {year} {2021})}\BibitemShut {NoStop}%
\bibitem [{\citenamefont {Cai}\ \emph {et~al.}(2019)\citenamefont {Cai},
  \citenamefont {Song}, \citenamefont {Wilson}, \citenamefont {Clark},
  \citenamefont {He}, \citenamefont {Zhang}, \citenamefont {Taniguchi},
  \citenamefont {Watanabe}, \citenamefont {Yao}, \citenamefont {Xiao},
  \citenamefont {McGuire}, \citenamefont {Cobden},\ and\ \citenamefont
  {Xu}}]{Cai2019}%
  \BibitemOpen
  \bibfield  {author} {\bibinfo {author} {\bibfnamefont {X.}~\bibnamefont
  {Cai}}, \bibinfo {author} {\bibfnamefont {T.}~\bibnamefont {Song}}, \bibinfo
  {author} {\bibfnamefont {N.~P.}\ \bibnamefont {Wilson}}, \bibinfo {author}
  {\bibfnamefont {G.}~\bibnamefont {Clark}}, \bibinfo {author} {\bibfnamefont
  {M.}~\bibnamefont {He}}, \bibinfo {author} {\bibfnamefont {X.}~\bibnamefont
  {Zhang}}, \bibinfo {author} {\bibfnamefont {T.}~\bibnamefont {Taniguchi}},
  \bibinfo {author} {\bibfnamefont {K.}~\bibnamefont {Watanabe}}, \bibinfo
  {author} {\bibfnamefont {W.}~\bibnamefont {Yao}}, \bibinfo {author}
  {\bibfnamefont {D.}~\bibnamefont {Xiao}}, \bibinfo {author} {\bibfnamefont
  {M.~A.}\ \bibnamefont {McGuire}}, \bibinfo {author} {\bibfnamefont {D.~H.}\
  \bibnamefont {Cobden}}, \ and\ \bibinfo {author} {\bibfnamefont
  {X.}~\bibnamefont {Xu}},\ }\href {\doibase 10.1021/acs.nanolett.9b01317}
  {\bibfield  {journal} {\bibinfo  {journal} {Nano Letters}\ }\textbf {\bibinfo
  {volume} {19}},\ \bibinfo {pages} {3993} (\bibinfo {year}
  {2019})}\BibitemShut {NoStop}%
\bibitem [{\citenamefont {Kim}\ \emph {et~al.}(2019)\citenamefont {Kim},
  \citenamefont {Yang}, \citenamefont {Li}, \citenamefont {Jiang},
  \citenamefont {Jin}, \citenamefont {Tao}, \citenamefont {Nichols},
  \citenamefont {Sfigakis}, \citenamefont {Zhong}, \citenamefont {Li},
  \citenamefont {Tian}, \citenamefont {Cory}, \citenamefont {Miao},
  \citenamefont {Shan}, \citenamefont {Mak}, \citenamefont {Lei}, \citenamefont
  {Sun}, \citenamefont {Zhao},\ and\ \citenamefont {Tsen}}]{Kim2019}%
  \BibitemOpen
  \bibfield  {author} {\bibinfo {author} {\bibfnamefont {H.~H.}\ \bibnamefont
  {Kim}}, \bibinfo {author} {\bibfnamefont {B.}~\bibnamefont {Yang}}, \bibinfo
  {author} {\bibfnamefont {S.}~\bibnamefont {Li}}, \bibinfo {author}
  {\bibfnamefont {S.}~\bibnamefont {Jiang}}, \bibinfo {author} {\bibfnamefont
  {C.}~\bibnamefont {Jin}}, \bibinfo {author} {\bibfnamefont {Z.}~\bibnamefont
  {Tao}}, \bibinfo {author} {\bibfnamefont {G.}~\bibnamefont {Nichols}},
  \bibinfo {author} {\bibfnamefont {F.}~\bibnamefont {Sfigakis}}, \bibinfo
  {author} {\bibfnamefont {S.}~\bibnamefont {Zhong}}, \bibinfo {author}
  {\bibfnamefont {C.}~\bibnamefont {Li}}, \bibinfo {author} {\bibfnamefont
  {S.}~\bibnamefont {Tian}}, \bibinfo {author} {\bibfnamefont {D.~G.}\
  \bibnamefont {Cory}}, \bibinfo {author} {\bibfnamefont {G.-X.}\ \bibnamefont
  {Miao}}, \bibinfo {author} {\bibfnamefont {J.}~\bibnamefont {Shan}}, \bibinfo
  {author} {\bibfnamefont {K.~F.}\ \bibnamefont {Mak}}, \bibinfo {author}
  {\bibfnamefont {H.}~\bibnamefont {Lei}}, \bibinfo {author} {\bibfnamefont
  {K.}~\bibnamefont {Sun}}, \bibinfo {author} {\bibfnamefont {L.}~\bibnamefont
  {Zhao}}, \ and\ \bibinfo {author} {\bibfnamefont {A.~W.}\ \bibnamefont
  {Tsen}},\ }\href {\doibase 10.1073/pnas.1902100116} {\bibfield  {journal}
  {\bibinfo  {journal} {Proceedings of the National Academy of Sciences}\
  }\textbf {\bibinfo {volume} {116}},\ \bibinfo {pages} {11131} (\bibinfo
  {year} {2019})}\BibitemShut {NoStop}%
\bibitem [{\citenamefont {Maleev}(1976)}]{maleev1976}%
  \BibitemOpen
  \bibfield  {author} {\bibinfo {author} {\bibfnamefont {S.~V.}\ \bibnamefont
  {Maleev}},\ }\href {http://www.jetp.ras.ru/cgi-bin/dn/e_043_06_1240.pdf}
  {\bibfield  {journal} {\bibinfo  {journal} {Sov. Phys. JETP}\ }\textbf
  {\bibinfo {volume} {43}},\ \bibinfo {pages} {1240} (\bibinfo {year}
  {1976})}\BibitemShut {NoStop}%
\bibitem [{\citenamefont {Kleinert}(1989)}]{kleinert_book}%
  \BibitemOpen
  \bibfield  {author} {\bibinfo {author} {\bibfnamefont {H.}~\bibnamefont
  {Kleinert}},\ }\href@noop {} {\emph {\bibinfo {title} {{Gauge Fields in
  Condensed Matter, Vol.1. Superlow and Vortex Lines}}}}\ (\bibinfo
  {publisher} {World Scientific, Singapore},\ \bibinfo {year}
  {1989})\BibitemShut {NoStop}%
\bibitem [{\citenamefont {Irkhin}\ and\ \citenamefont
  {Katanin}(1999)}]{PhysRevB.60.2990}%
  \BibitemOpen
  \bibfield  {author} {\bibinfo {author} {\bibfnamefont {V.~Y.}\ \bibnamefont
  {Irkhin}}\ and\ \bibinfo {author} {\bibfnamefont {A.~A.}\ \bibnamefont
  {Katanin}},\ }\href {\doibase 10.1103/PhysRevB.60.2990} {\bibfield  {journal}
  {\bibinfo  {journal} {Phys. Rev. B}\ }\textbf {\bibinfo {volume} {60}},\
  \bibinfo {pages} {2990} (\bibinfo {year} {1999})}\BibitemShut {NoStop}%
\bibitem [{\citenamefont {Grechnev}\ \emph {et~al.}(2005)\citenamefont
  {Grechnev}, \citenamefont {Irkhin}, \citenamefont {Katsnelson},\ and\
  \citenamefont {Eriksson}}]{PhysRevB.71.024427}%
  \BibitemOpen
  \bibfield  {author} {\bibinfo {author} {\bibfnamefont {A.}~\bibnamefont
  {Grechnev}}, \bibinfo {author} {\bibfnamefont {V.~Y.}\ \bibnamefont
  {Irkhin}}, \bibinfo {author} {\bibfnamefont {M.~I.}\ \bibnamefont
  {Katsnelson}}, \ and\ \bibinfo {author} {\bibfnamefont {O.}~\bibnamefont
  {Eriksson}},\ }\href {\doibase 10.1103/PhysRevB.71.024427} {\bibfield
  {journal} {\bibinfo  {journal} {Phys. Rev. B}\ }\textbf {\bibinfo {volume}
  {71}},\ \bibinfo {pages} {024427} (\bibinfo {year} {2005})}\BibitemShut
  {NoStop}%
\bibitem [{\citenamefont {Bolvin}(2006)}]{Bolvin2006}%
  \BibitemOpen
  \bibfield  {author} {\bibinfo {author} {\bibfnamefont {H.}~\bibnamefont
  {Bolvin}},\ }\href {\doibase {10.1002/cphc.200600051}} {\bibfield  {journal}
  {\bibinfo  {journal} {ChemPhysChem}\ }\textbf {\bibinfo {volume} {7}},\
  \bibinfo {eid} {1575} (\bibinfo {year} {2006})}\BibitemShut {NoStop}%
\bibitem [{\citenamefont {Goodenough}(1958)}]{Goodenough1958}%
  \BibitemOpen
  \bibfield  {author} {\bibinfo {author} {\bibfnamefont {J.~B.}\ \bibnamefont
  {Goodenough}},\ }\href {\doibase
  https://doi.org/10.1016/0022-3697(58)90107-0} {\bibfield  {journal} {\bibinfo
   {journal} {Journal of Physics and Chemistry of Solids}\ }\textbf {\bibinfo
  {volume} {6}},\ \bibinfo {pages} {287} (\bibinfo {year} {1958})}\BibitemShut
  {NoStop}%
\bibitem [{\citenamefont {Kanamori}(1959)}]{Kanamori1959}%
  \BibitemOpen
  \bibfield  {author} {\bibinfo {author} {\bibfnamefont {J.}~\bibnamefont
  {Kanamori}},\ }\href {\doibase https://doi.org/10.1016/0022-3697(59)90061-7}
  {\bibfield  {journal} {\bibinfo  {journal} {Journal of Physics and Chemistry
  of Solids}\ }\textbf {\bibinfo {volume} {10}},\ \bibinfo {pages} {87}
  (\bibinfo {year} {1959})}\BibitemShut {NoStop}%
\bibitem [{\citenamefont {Ni}\ \emph {et~al.}(2021)\citenamefont {Ni},
  \citenamefont {Li}, \citenamefont {Amoroso}, \citenamefont {He},
  \citenamefont {Feng}, \citenamefont {Kan}, \citenamefont {Picozzi},\ and\
  \citenamefont {Xiang}}]{Ni2021}%
  \BibitemOpen
  \bibfield  {author} {\bibinfo {author} {\bibfnamefont {J.~Y.}\ \bibnamefont
  {Ni}}, \bibinfo {author} {\bibfnamefont {X.~Y.}\ \bibnamefont {Li}}, \bibinfo
  {author} {\bibfnamefont {D.}~\bibnamefont {Amoroso}}, \bibinfo {author}
  {\bibfnamefont {X.}~\bibnamefont {He}}, \bibinfo {author} {\bibfnamefont
  {J.~S.}\ \bibnamefont {Feng}}, \bibinfo {author} {\bibfnamefont {E.~J.}\
  \bibnamefont {Kan}}, \bibinfo {author} {\bibfnamefont {S.}~\bibnamefont
  {Picozzi}}, \ and\ \bibinfo {author} {\bibfnamefont {H.~J.}\ \bibnamefont
  {Xiang}},\ }\href {\doibase 10.1103/PhysRevLett.127.247204} {\bibfield
  {journal} {\bibinfo  {journal} {Physical Review Letters}\ }\textbf {\bibinfo
  {volume} {127}},\ \bibinfo {pages} {247204} (\bibinfo {year}
  {2021})}\BibitemShut {NoStop}%
\bibitem [{\citenamefont {Torelli}\ and\ \citenamefont
  {Olsen}(2018)}]{Torelli_2018}%
  \BibitemOpen
  \bibfield  {author} {\bibinfo {author} {\bibfnamefont {D.}~\bibnamefont
  {Torelli}}\ and\ \bibinfo {author} {\bibfnamefont {T.}~\bibnamefont
  {Olsen}},\ }\href {\doibase 10.1088/2053-1583/aaf06d} {\bibfield  {journal}
  {\bibinfo  {journal} {2D Materials}\ }\textbf {\bibinfo {volume} {6}},\
  \bibinfo {pages} {015028} (\bibinfo {year} {2018})}\BibitemShut {NoStop}%
\bibitem [{\citenamefont {Xu}\ \emph {et~al.}(2018)\citenamefont {Xu},
  \citenamefont {Feng}, \citenamefont {Xiang},\ and\ \citenamefont
  {Bellaiche}}]{Xu2018}%
  \BibitemOpen
  \bibfield  {author} {\bibinfo {author} {\bibfnamefont {C.}~\bibnamefont
  {Xu}}, \bibinfo {author} {\bibfnamefont {J.}~\bibnamefont {Feng}}, \bibinfo
  {author} {\bibfnamefont {H.}~\bibnamefont {Xiang}}, \ and\ \bibinfo {author}
  {\bibfnamefont {L.}~\bibnamefont {Bellaiche}},\ }\href {\doibase
  10.1038/s41524-018-0115-6} {\bibfield  {journal} {\bibinfo  {journal} {npj
  Computational Materials}\ }\textbf {\bibinfo {volume} {4}},\ \bibinfo {pages}
  {57} (\bibinfo {year} {2018})}\BibitemShut {NoStop}%
\end{thebibliography}%
 
\clearpage
\newpage

\begin{figure*}[h!]
    \centering
    \includegraphics[width=1\columnwidth]{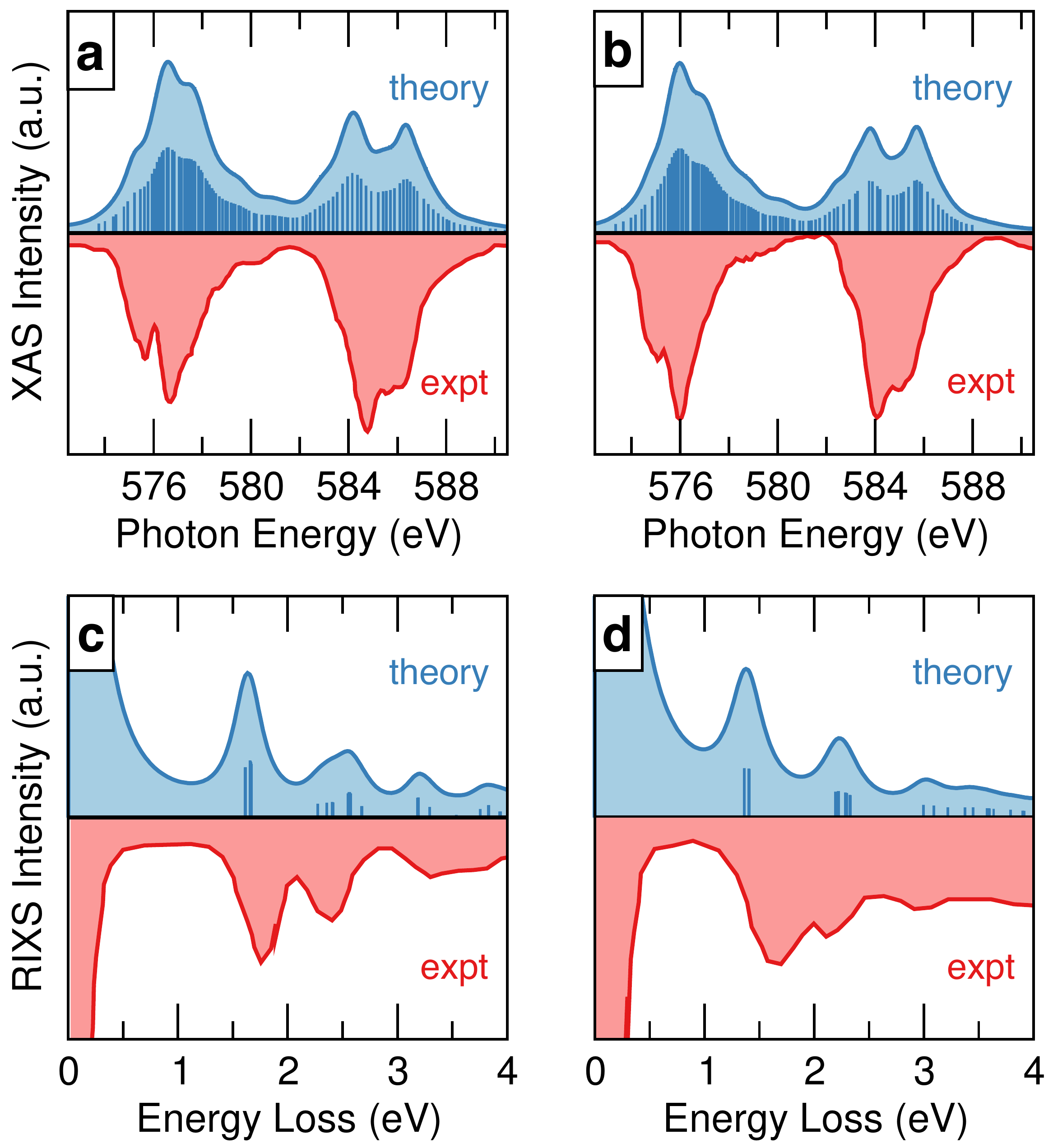}
    \caption{Simulated and experimental x-ray absorption spectra (XAS) of (a) CrCl$_3$ and (b) CrI$_3$. Simulated and experimental resonant x-ray inelastic scattering (RIXS) spectra of (c) CrCl$_3$ and (d) CrI$_3$. Vertical lines in the simulated spectra mark the excitation energies. The experimental spectra are digitized from Ref.\ \cite{Shao2021}. The simulated XAS spectra were shifted to match the position in energy of the highest intensity peak in the experimental spectra. Notice the different order of intensity between theory and experiment observed for the $L_3$- and $L_2$-edges is due to undesired self-absorption effects in experimental measurements. \label{Fig2}}
\end{figure*}

\begin{table*}
\caption{Relative energies of the $3d^3$ multiplet structure of Cr$^{3+}$  ions in Cr$X_3$ ($X =$ Cl, Br, I), as obtained from CASSCF and MRCI calculations. Energies are given in eV and referenced to the ground state. Occupations in parentheses indicate the dominant configuration.}
\centering
\ra{1.3}
\begin{tabular}{@{}lccccccccccccccc@{}}\toprule
&  \multicolumn{2}{c}{CrCl$_3$} & \phantom{ab}& \multicolumn{2}{c}{CrBr$_3$}  \phantom{ab}& \multicolumn{2}{c}{CrI$_3$} \\ \cmidrule(lr){2-3} \cmidrule(lr){5-6} \cmidrule(lr){8-9}
& CASSCF & MRCI && CASSCF & MRCI && CASSCF & MRCI \\ \midrule
$^4A_2$ ($t^3_{2g}e^0_{g}$) & 	0.00			 	& 0.00		    	 && 	0.00			 	& 	0.00	&& 	0.00			 	& 	0.00 \\
$^4T_2$ ($t^2_{2g}e^1_{g}$) & 	1.57, 1.58, 1.61 	& 1.67, 1.68, 1.72  	&& 	1.38, 1.59, 2.00 	& 	1.46, 1.66, 2.05 && 	1.44, 1.56, 1.58 	& 	1.48, 1.59, 1.61\\
$^2E$ ($t^3_{2g}e^0_{g}$) & 	2.36, 2.36 		& 2.22, 2.22  		&& 	1.92, 2.19 		& 	1.77, 2.10  && 	2.32, 2.33 		& 	2.16, 2.17 \\
$^4T_1$ ($t^2_{2g}e^1_{g}$) & 	2.51, 2.53, 2.60 	& 2.50, 2.52, 2.59  	&& 	2.51, 2.59, 2.95 	& 	2.49, 2.62, 2.96 && 	2.48, 2.50, 2.52 	& 	2.39, 2.49, 2.51 \\
$^2T_1$ ($t^3_{2g}e^0_{g}$) & 	2.45, 2.47, 2.48 	& 2.31, 2.33, 2.34  	&& 	2.36, 2.45, 2.49 	& 	2.22, 2.32, 2.37 && 	2.39, 2.39, 2.42 	& 	2.26, 2.30, 2.31 \\
$^2T_1$ ($t^2_{2g}e^1_{g}$) & 	3.30, 3.31, 3.33 	& 3.03, 3.05  		&& 	3.32, 3.41, 3.45 	& 	3.23, 3.25, 3.26 && 	3.23, 3.29, 3.31 	& 	2.95, 3.02, 3.02 \\
$^2A_1$ ($t^2_{2g}e^1_{g}$) & 	3.56 				& 3.47 			 && 	3.56 				& 	3.45 && 	3.47 				& 	3.39 \\
$^2T_1$ ($t^2_{2g}e^1_{g}$) & 	3.83, 3.84, 3.85 	& 3.76, 3.77, 3.78  	&& 	4.06, 4.32, 4.80 	& 	3.94, 4.24, 4.73 && 	3.68, 3.70, 3.71 	& 	3.59, 3.61, 3.63 \\
$^4T_1$ ($t^1_{2g}e^2_{g}$) & 	4.13, 4.15, 4.16 	& 4.05, 4.07, 4.08  	&& 	3.71, 3.85, 3.89 	& 	3.56, 3.74, 3.78 && 	3.90, 4.01, 4.04 	& 	3.80, 4.10, 4.12 \\ \bottomrule
\end{tabular}
\label{Table1}
\end{table*}

\begin{table*}
\caption{Intra-site magnetic interactions in Cr$X_3$ ($X =$ Cl, Br, I), i.e., single-ion anisotropy axial parameter $A_{\rm sia}$, along with the in-plane ($g_{xx}$, $g_{yy}$) and out-of-plane ($g_{zz}$) components of the $g$-tensor, as obtained from  CASSCF and MRCI calculations.}
\centering
\ra{1.3}
\begin{tabular}{@{}lccccccccccccc@{}}\toprule
&  \multicolumn{2}{c}{CrCl$_3$} & \phantom{ab}& \multicolumn{2}{c}{CrBr$_3$} &
\phantom{ab} & \phantom{ab}& \multicolumn{2}{c}{CrI$_3$} &
\phantom{ab} \\ \cmidrule(lr){2-3} \cmidrule(lr){5-6} \cmidrule(lr){8-9}
									& CASSCF 			& MRCI 				&& CASSCF 				& MRCI  && CASSCF 				& MRCI\\ \midrule
$A_{\rm sia}$ (meV) 							& $-0.02$ 			& $-0.03$    			&&   $-0.07$ 				& $-0.08$ &&   $-0.11$ 				& $-0.12$ \\ 
$g_{xx} = g_{yy}$  						& $1.44$ 				& $1.45$    			&&   $1.86$ 				& $1.86$ &&   $1.91$ 				& $1.92$\\ 
$g_{zz}$  								& $1.46$ 				& $1.47$   			&&   $1.87$ 				& $1.88$ &&   $1.92$ 				& $1.93$\\\bottomrule
\end{tabular}
\label{Table2}
\end{table*}

\begin{table*}
\caption{Comparison of the dipolar anisotropy parameter $A_{\rm dip}$ and the single-ion anisotropy parameter $A_{\rm sia}$ obtained from MRCI calculations for bulk Cr$X_3$ ($X =$ Cl, Br, I). The negative sign indicates that the in-plane spin moments are favoured. The shape anisotropy parameter $A_{\rm dip}$ for monolayer Cr$X_3$ ($X =$ Cl, Br, I) approximately doubles in magnitude as compared to the bulk, and attains values of 144~$\mu$eV, 131~$\mu$eV, 108~$\mu$eV for chloride, bromide and iodide, respectively.}
\centering
\ra{1.3}
\begin{tabular}{crrrrr}\toprule
\phantom{abcde} &  CrCl$_3$ & \phantom{ab} & CrBr$_3$ &\phantom{ab}& CrI$_3$  \\
						 \midrule
$A_{\rm sia}$  ($\mu$eV)		& $-31.4$  & \phantom{ab} & 	$-81.1$ & \phantom{ab} 	& $-124.2$ \\ 
$A_{\rm dip}$  ($\mu$eV)		& $85.2$  & \phantom{ab} & 	$75.5$ & \phantom{ab} 	& $57.3$ \\  \bottomrule
\end{tabular}
\label{shape_aniso}
\end{table*}

\begin{table*}
\caption{Inter-site magnetic interactions in Cr$X_3$ ($X =$ Cl, Br, I), {\it i.e.} bilinear ($J_1$) and biquadratic ($J_2$) exchange couplings, along with the off-diagonal ($\Gamma_{xy}$, $\Gamma_{yz}$, $\Gamma_{zx}$) and Kitaev ($K$) parameters comprised in the symmetric anisotropic tensors, as obtained from the CASSCF and MRCI calculations.}
\centering
\ra{1.3}
\begin{tabular}{@{}lcccccccccccccc@{}}\toprule
&  \multicolumn{2}{c}{CrCl$_3$} & \phantom{ab}& \multicolumn{2}{c}{CrBr$_3$} &
\phantom{ab}& \multicolumn{2}{c}{CrI$_3$} & \phantom{ab}  \\  \cmidrule(lr){2-3}  \cmidrule(lr){5-6}  \cmidrule(lr){8-9}
									& CASSCF 			& MRCI 				&& CASSCF 				& MRCI && CASSCF 				& MRCI \\ \midrule
$J_1$ (meV)  							& $-0.64$ 				& $-0.97$   			&&   $-0.61$ 				& $-1.21$ &&   $-0.60$ 				& $-1.38$ \\ 
$J_2$ (meV)  							& $-0.04 $			& $-0.05$    			&&   $-0.05$ 				& $-0.05$ &&   $-0.06$ 				& $-0.06$\\ 
$\Gamma_{xy}$ (meV) 					& $-1.2 \times 10^{-4}$ 	& $-2.1 \times 10^{-4}$    	&&   $-6.9 \times 10^{-4}$ 		& $-0.8 \times 10^{-3}$ &&   $-1.2 \times 10^{-3}$ 		& $-4.2 \times 10^{-4}$\\ 
$\Gamma_{yz} = -\Gamma_{zx}$  (meV)		& $-1.3 \times 10^{-5}$ 	& $-0.8 \times 10^{-4}$    	&&   $-6.9 \times 10^{-4}$ 		& $-0.9 \times 10^{-3}$ &&   $-1.1 \times 10^{-4}$ 		& $-3.1 \times 10^{-4}$\\ 
$K$ (meV) 							& $-1.7 \times 10^{-4}$ 	& $-1.1 \times 10^{-4}$    	&&   $-8.2 \times 10^{-3}$ 		& $-0.01$ &&   $-9.3 \times 10^{-3}$ 		& $-0.05$\\\bottomrule
\end{tabular}
\label{Table3}
\end{table*}

\end{document}